\documentclass[a4paper,11pt]{article}
\pdfoutput=1
\usepackage{jcappub}
\usepackage{txfonts}
\usepackage{threeparttable}
\usepackage{aas_macros}
\usepackage{graphicx}
\usepackage{subfigure}
\usepackage{url}
\usepackage{amsmath}

\def\qm{q_\mathrm{m}}

\title{Precision cosmology with time delay lenses: high resolution imaging requirements}

\author[1,2]{Xiao-Lei Meng,}
\author[2,3]{Tommaso Treu,}
\author[2,3]{Adriano Agnello,}
\author[4]{Matthew W.~Auger,}
\author[1,2,3]{Kai Liao,}
\author[5]{Philip J. Marshall}

\affiliation[1]{Department of Astronomy, Beijing Normal University, Beijing, 100875, China}
\affiliation[2]{Department of Physics, University of California, Santa Barbara, CA 93106, USA}
\affiliation[3]{Physics and Astronomy Department, University of California, Los Angeles, CA 90095-1547, USA}
\affiliation[4]{Institute of Astronomy, University of Cambridge, Madingley Road, Cambridge CB3 0HA, UK}
\affiliation[5]{Kavli Institute for Particle Astrophysics and Cosmology, Stanford University, 452 Lomita Mall, Stanford, CA 94305, USA}

\emailAdd{xlmeng919@gmail.com}

\date{Accepted . Received }

\abstract{ Lens time delays are a powerful probe of cosmology,
provided that the gravitational potential of the main deflector can be
modeled with sufficient precision. Recent work has shown that this can
be achieved by detailed modeling of the host galaxies of lensed
quasars, which appear as ``Einstein Rings'' in high resolution images.
The distortion of these arcs and counter-arcs, as measured over a
large number of pixels, provides tight constraints on the difference
between the gravitational potential between the quasar image
positions, and thus on cosmology in combination with the measured time
delay. We carry out a systematic exploration of the high resolution
imaging required to exploit the thousands of lensed quasars that will
be discovered by current and upcoming surveys with the next decade.
Specifically, we simulate realistic lens systems as imaged by the
Hubble Space Telescope (HST), James Webb Space Telescope (JWST), and
ground based adaptive optics images taken with Keck or the Thirty
Meter Telescope (TMT). We compare the performance of these pointed
observations with that of images taken by the Euclid (VIS), Wide-Field
Infrared Survey Telescope (WFIRST) and Large Synoptic Survey Telescope
(LSST) surveys. We use as our metric the precision with which the
slope $\gamma'$ of the total mass density profile $\rho_{tot}\propto
r^{-\gamma'}$ for the main deflector can be measured. Ideally, we
require that the statistical error on $\gamma'$ be less than 0.02,
such that it is subdominant to other sources of random and systematic
uncertainties. We find that survey data will likely have sufficient
depth and resolution to meet the target only for the brighter
gravitational lens systems, comparable to those discovered by the SDSS
survey. For fainter systems, that will be discovered by current and
future surveys, targeted follow-up will be required. However, the
exposure time required with upcoming facilitites such as JWST, the
Keck Next Generation Adaptive Optics System, and TMT, will only be of
order a few minutes per system, thus making the follow-up of hundreds
of systems a practical and efficient cosmological probe.}


\begin{document}
\newcommand\farcs{\hbox{$.\!\!^{\prime\prime}$}}
\maketitle
\flushbottom


\section{Introduction}

In the past few years, gravitational time delays (Refsdal 1964
\citep{1964MNRAS.128..307R}) have emerged as a powerful and cost
effective cosmological probe. Studies based on blind analysis have
shown that a single system consisting of multiple images of a
background quasar (at redshift $z_{\rm s}$) lensed by a foreground
massive elliptical galaxy at redshift $z_{\rm d}$ can be used to
measure the so-called time delay distance with an uncertainty of 5-7\%
(Suyu et al. 2010 \citep{2010ApJ...711..201S}, 2013
\citep{2013ApJ...766...70S}, 2014 \citep{2014ApJ...788L..35S}). The
time delay distance $D_{\Delta t}$ gives a direct measurement of the
Hubble constant and allows one to break some of the main degeneracies
in the interpretation of cosmic microwave background data, thus
providing tight constraints on parameters such as curvature and dark
energy equation of state (Coe \& Moustakas 2009
\citep{2009ApJ...706...45C}, Linder 2011 \citep{2011PhRvD..84l3529L},
Weinberg 2013 \citep{2013PhR...530...87W}, Suyu 2012
\citep{2012MNRAS.426..868S}, Suyu et al. 2014
\citep{2014ApJ...788L..35S}).  The time delay distance
measurement of H$_0$ is comparable in terms of information content to
that obtained via the cosmic distance ladder (Riess et al. 2011
\citep{2011ApJ...730..119R}, Freedman et al. 2012
\citep{2012ApJ...758...24F}), in that is based on absolute distances
and completely independent of the properties of the early universe
($z>z_{\rm s}$). Importantly, time delay distances are independent of
the {\it local} distance ladder, and thus provide a crucial test of
any potential systematic uncertainties. Furthermore, being
independent, the cosmic distance ladder and time delay distance
constraints on H$_0$ can be statistically combined for additional
gains in precision.

From an observational point of view, the attainment of precise
(i.e. small random errors) and accurate (small systematic errors) time
delay measurements relies on a few important ingredients. First,
monitoring of the lensed quasars is required to obtain time delays
with a few percent uncertainties. Dedicated monitoring campaigns have
shown that this precision is achievable with 1-2m class telescopes at
present time (e.g. from COMOGRAIL; Tewes et al. 2013
\citep{2013A&A...556A..22T}), or in the radio (Fassnacht et al. 2002
\citep{2002ApJ...581..823F}). In the future, the exploitation of
larger samples of lenses (Oguri \& Marshall 2010
\cite{2010MNRAS.405.2579O}) will require monitoring campaigns on 2-4m
telescopes (Treu et al. 2013 \citep{2013arXiv1306.1272T}), or the
deployment of long term high-cadence monitoring efforts, like the LSST
(Liao et al. 2015 \citep{2015ApJ...800...11L}). Second, the
spectroscopic redshift of the source and deflector must be measured.
For current samples, this is typically a relatively straightforward
step, requiring short exposures on 4-10m class telescopes (Fassnacht \&
Cohen 1998 \citep{1998AJ....115..377F}, Eigenbrod et al. 2007
\citep{2007A&A...465...51E}).  Third, the gravitational potential of
the main deflector needs to be constrained by the data so that the
uncertainty on its difference between the location of the images is
also of order 3\%. This goal can be achieved by studying the extended
structure of the lensed quasar host galaxy, and the stellar kinematics
of the deflector galaxy (Kochanek et al. 2001
\citep{2001ApJ...547...50K}, Treu \& Koopmans 2002
\citep{2002MNRAS.337L...6T}, Koopmans et al. 2003
\citep{2003ApJ...599...70K}, Suyu et al. 2010
\citep{2010ApJ...711..201S}, Suyu et al. 2013
\citep{2013ApJ...766...70S}, Suyu et al. 2014
\cite{2014ApJ...788L..35S}).  Fourth, the combined effects of the
inhomogenous mass distribution along the line of sight need to be
taken into account. Recent work has shown that the line of sight
effects can be sufficiently characterized by measuring the properties
of galaxies and weak lensing signal in the field of the main
deflector, and comparing with simulated lines of sight (Suyu et al.
2010 \citep{2010ApJ...711..201S}, 2013 \citep{2013ApJ...766...70S},
2014 \citep{2014ApJ...788L..35S}, Greene et al. 2013
\citep{2013ApJ...768...39G}, Collett et al. 2013
\citep{2013MNRAS.432..679C}). Fifth, the mass sheet degeneracy (Falco et 
al. 1985 \cite{FGS85}) and its generalizations (Schneider \& Sluse
2013, 2014, Xu et al. 2015; \cite{SS13,SS14,Xu15}) must be broken
either by the incorporation of non-lensing data such as stellar
velocity dispersion (Treu \& Koopmans 2002, Koopmans et al. 2003, Suyu
et al. 2014,
\cite{2002MNRAS.337L...6T,2003ApJ...599...70K,2014ApJ...788L..35S}),
or by appropriate asssumptions about the mass distribution in the main
deflector (Xu et al. 2015 \citep{Xu15}).
  
However, whereas current samples have been limited to a few lenses,
current and ongoing surveys---such as the Dark Energy Survey (DES),
PanSTARRS1, HyperSuprimeCam (HSC)--- should together discover hundreds
of lensed quasars suitable for monitoring and follow-up (Oguri \&
Marshall 2010 \cite{2010MNRAS.405.2579O}). During the next decade, the
Large Synoptic Survey Telescope (LSST), Euclid, and the Wide-Field
Infrared Survey Telescope (WFIRST) will come online, and enable the
detection and measurement of an order of magnitude more, with LSST
likely to provide high accuracy time delays for several hundred
systems (Liao et al. 2015 \cite{2015ApJ...800...11L}). These large
samples will be extremely powerful cosmographic probes (Coe \&
Moustakas 2009 \citep{2009ApJ...706...45C}, Linder 2011
\citep{2011PhRvD..84l3529L}), provided that sufficient follow-up data
are available (Treu et al. 2013 \citep{2013arXiv1306.1272T}). In this
paper we carry out a feasibility study for cosmology with future time
delay lenses, focusing on the high-resolution imaging follow-up
requirements. The multi-pronged nature of time delay lens follow-up
makes it natural to follow the approach of addressing each component
of the follow-up independently. The monitoring and spectroscopic
follow-up requirements are described, for example, by Eigenbrod et al.
(2005; \cite{2005A&A...436...25E}) and Linder (2015;
\cite{2015PhRvD..91h3511L}). Specifically, we aim to answer the
following questions:
\begin{itemize}
    \item Will sufficient information be available from the survey
            data itself, or will dedicated follow-up be necessary?
    \item If dedicated follow-up is necessary, approximately how much
            exposure time will be required per system, to enable
            high precision cosmography?
\end{itemize}
In order to make the problem tractable and the results general, we
need a single simple metric to evaluate the quality of imaging data.
In short, we need to quantify our ability to measure how the
deflection angle (i.e. the derivative of the lensing potential) varies
between the images, and thus how the extended images of the quasar
host galaxy are stretched across the image. In practice, our ability
to constrain the differential magnification will depend on the
resolution of the data as well as on the number of pixels where the
source is detected above a certain signal to noise ratio (and thus on
the source magnification for fixed instrumental configuration and
source intrinsic luminosity).

We choose to adopt as our metric the slope $\gamma'$ of the total mass
density profile of the form $\rho_{\rm tot}\propto r^{-\gamma'}$ as
measured by fitting elliptical power law models to the data. This
profile is the simplest one that provides a realistic description of
galaxy scale lenses (Treu 2010 \citep{2010ARA&A..48...87T}), and the
uncertainty $\gamma'$ has been shown to be approximately proportional
to the uncertainty on the gravitational potential differences and thus
the time delay distance, and the Hubble constant (Kochanek 2002
\citep{2002ApJ...578...25K}, Wucknitz 2002
\citep{2002MNRAS.332..951W}, Suyu 2012
\citep{2012MNRAS.426..868S}).  Intuitively, $\gamma'$ is directly
related to the variations of the lensing potential. In fact, if
$\gamma'=2$ (the so-called isothermal profile) the deflection angle is
constant across the image and therefore all the images will appear to
have the same radial magnification. If the profile is
steeper/shallower, radial magnification will vary across the image,
thus giving rise to images of different widths.

Naturally, for a real measurement it is important to explore different
profiles and the role of the choice of the profile in the
uncertainties (see e.g. Suyu et al 2014 and references therein
\cite{2014ApJ...788L..35S}). However, our goal is to estimate minimal
requirements on the data quality. If the data quality is insufficient
to constrain $\gamma'$ it will also be insufficient to constrain more
flexible models. Likewise, this paper is only concerned with
precision, i.e. random errors. Systematic errors not considered in
this paper, such as those arising from incomplete knowledge of the PSF
will only increase the error budget (e.g., Agnello et al. 2015; Rusu
et al. 2015;
\citep{2015arXiv150602720A,2015arXiv150605147R}).
%
Thus, in order to leave room for additional sources of uncertainty in
the error budget, we set a rather stringent requirement of 0.02
uncertainty of $\gamma'$ (corresponding approximately to 2\% per
system on time delay distance), in tests where the mock data are
generated using the same model as is used in the inference of the lens
parameters. In other words, our target corresponds to the requirement
that the {\it statistical} error arising from the image quality be
subdominant with respect to those arising from modeling uncertainties,
time delays, and line of sight effects.

This paper is organized as follows. In Section 2, we summarize the
characteristics of the telescopes and instruments simulated as part of
this work. Then, in Section 3, we describe the properties of the
simulated lenses. Next, in Section 4 we describe the procedure used to
carry out the inference.  Our results are presented in Section
5. Finally, we discuss and summarize our work in Section 6. Throughout
this paper, all magnitudes are given in the AB system. Even though our
findings are independent of any cosmology, we adopt a spatially flat
$\Lambda$CDM cosmology with $\Omega_{\rm{m}} = 0.3$, $\Omega_{\Lambda}
= 0.7$, and the Hubble constant $H_{0} = 70~\rm km~s^{-1}~Mpc^{-1}$
when calculating distances.


\section{Summary of simulated instrumental setups}

We aim to carry out a systematic exploration of imaging requirements
given current and future facilities. For a given lens system, the
instrumental setup drives our ability to constrain $\gamma'$ in a
number of ways. First, the signal-to-noise ratio depends on the
exposure-time adopted, sky or background noise in a chosen band, and
the instrumental readout noise. Second, the pixel-size and the
point-spread function (PSF) properties determine how finely one can
map the system being observed, hence how robustly the deflections on
either side of the lens can be quantified.

Here we summarize the main properties of instruments that will be
considered in our simulations. In an effort to balance completeness
with feasibility, we consider a suite of instruments that include
ground and space based telescopes, and cadenced surveys where the
total exposure cannot be set arbitrarily. Some of the choices are
representative of a class of telescopes/instruments, for example we
expect the performance of Keck Adaptive Optics to be a useful guidance
for adaptive optics systems on other 8-10m class telescopes, and that
of the Thirty Meter Telescope to be a guidance for other planned
20-30m class telescopes. For survey telescopes like WFIRST, LSST, and
Euclid, we only consider the wide field surveys, as the deep surveys
cover too small a solid angle to include randomly a useful number of
lensed quasars.

\begin{figure}
\begin{center}
\includegraphics[width=0.9\textwidth]{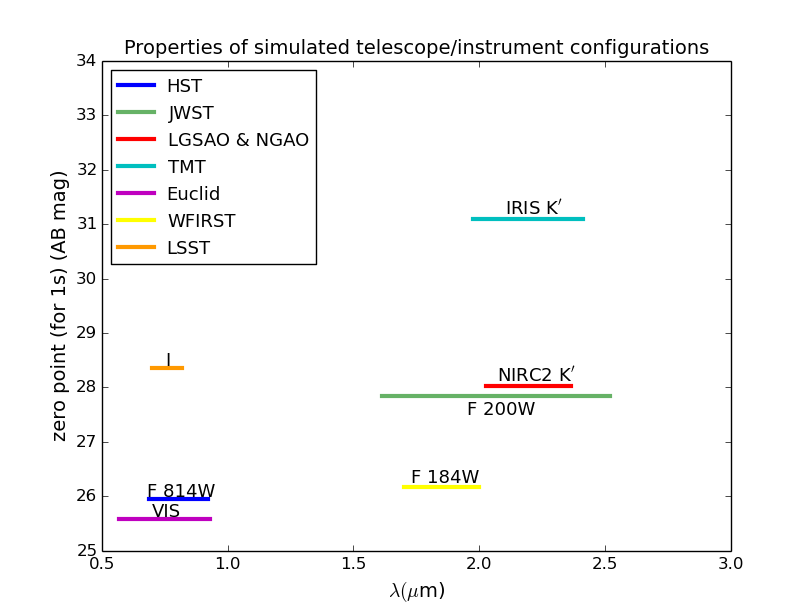}
\end{center}
\caption{Zero points in AB magnitudes for HST/ACS (blue), JWST/NIRCAM (green), Keck NIRC2  (assumed for both LGSAO and NGAO; red), TMT/IRIS (cyan), Euclid (magenta), WFIRST (yellow) and LSST (orange), corresponding to one count per second. The colored bars indicate the wavelength range of each instrumental setup considered in this work.}
\label{fig:zp_wavelength}
\end{figure}

\begin{figure}
\begin{center}
\includegraphics[width=1.0\textwidth]{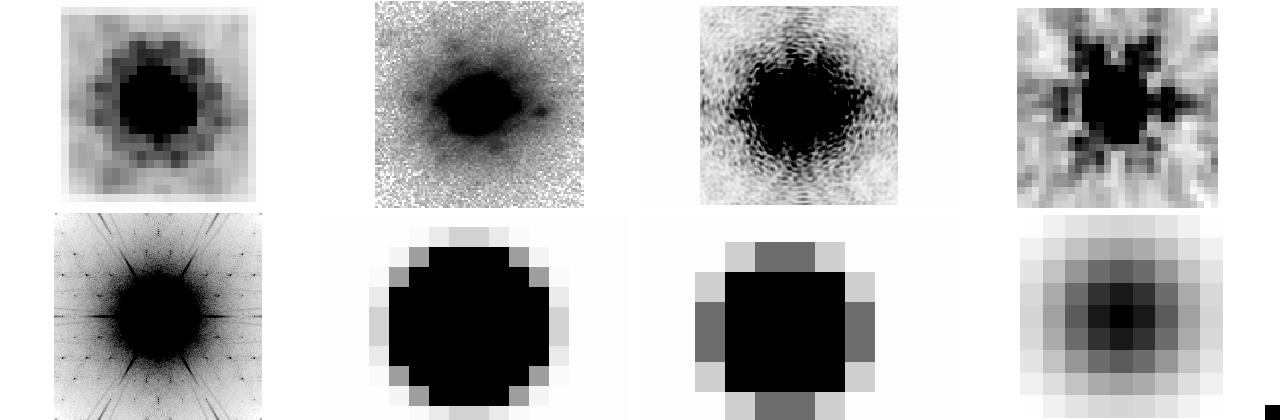}
\end{center}
\caption{Montage of the point spread functions (PSFs) of each instrument. The upper row, from left to right, represents HST/ACS, Keck/NIRC2+LGSAO, Keck/NIRC2+NGAO, and JWST/NIRCAM. The lower row, from left to right, shows TMT/IRIS, Euclid, WFIRST, and LSST, respectively. Observed or simulated PSFs are used for HST, JWST, Keck (LGSAO \& NGAO), and TMT. Gaussian PSFs are adopted for the other three survey instruments.}
\label{fig:PSF_montage}
\end{figure}

Table~\ref{tab:telescopes parameters} and
figure~\ref{fig:zp_wavelength} display the main properties of the
telescopes that are needed when generating images of mock
gravitational lenses, taken from the instrument
websites. Figure~\ref{fig:PSF_montage} shows the typical PSF of each
instrument. In order to facilitate comparisons between instruments and
telescopes, whenever possible we selected the filter/band for each
configuration that is closer to the $I$-band in the optical and the
$K'$-band in the near infrared \footnote{Clearly, for survey
telescopes such as LSST one can further increase the precision on
$\gamma'$ by considering multiple bands at once, when they are
available (Newton et al. 2011 \citep{2011ApJ...734..104N},
Auger et al. 2013 \citep{2013MNRAS.436..503A}). However, the
multifilter analysis is beyond the scope of this paper, which is
concerned with comparison across instruments/telescopes.}. A brief
description of each setup follows.

\begin{enumerate}

\item  We take the Advanced Camera for Surveys (ACS) as our reference
optical imager on board the Hubble Space Telescope (HST). Its
properties are taken from the Space Telescope Science Institute HST
Exposure Time Calculator.\footnote{\url{http://etc.stsci.edu}} In the
chosen F814W filter its performance is comparable to that of WFC3
(slightly higher sensitivity and coarser pixel size). For simplicity,
we neglect charge transfer inefficiency effects, assuming that they
are negligible or can be corrected to the desired level. The PSF is
simulated using the Tiny Tim software (Krist 1993
\cite{1993ASPC...52..536K}).

\item The Near Infrared Camera (NIRCAM) on board the James Webb Space
Telescope (JWST) is chosen as the next space based imaging capability.
We select the broad F200W filter, where the image quality is virtually
diffraction limited. We use the instrument properties as given by the
JWST Exposure Time Calculator.\footnote{\url{http://jwstetc.stsci.edu}}

\item The current and planned adaptive optics systems (hereafter LGSAO \&
NGAO;\citep{2006PASP..118..297W}, \citep{2010SPIE.7736E..0KW},
respectively) at the W.M.Keck Observatory are chosen to represent
current and upcoming AO performance on 8-10m class telescopes. We
consider the current instrument NIRC2, although further gains might be
possible with an instrument upgrade in conjuction with the AO system
upgrade. We also assume conservatively the same background for NGAO as
with LGSAO, even though the thermal background should be significantly
lower for NGAO, due to the lower operating temperature of the NGAO
system. We adopt the typical configuration used for studies of
gravitational lens systems (Fassnacht \& Cohen 1998
\citep{1998AJ....115..377F}): $K'$ filter, MCDS readout mode, read
number $N=16$, and narrow camera mode. A real observed PSF is used for
the current AO system. The simulated PSF and performance
characteristics of NGAO have been kindly provided by the NGAO team.

\item The infrared imager and spectrograph IRIS (Larkin et al. 2010
\citep{2010SPIE.7735E..29L}) working behind adaptive optics on the
Thirty Meter Telescope (TMT) is selected to represent the performance
of the extremely large telescopes that will be operational in the next
decade. IRIS is expected to be close to diffraction limited in the
$K'$ band. The simulated PSF and performance characteristics of
TMT-IRIS have been obtained from the IRIS team, from the TMT Exposure
Time Calculator (ETC)
\footnote{\url{http://tmt.mtk.nao.ac.jp/ETC_readme.html}} and from the
paper by Do et al. (2014) \citep{2014AJ....147...93D}.

\item Euclid is a space survey telescope planned to be launched by the
European Space Agency. Euclid has two instruments: a high-resolution
visible imager, and a NIR imaging spectrograph. We study the performance
of the high-resolution visible imager, the more suitable to detailed
gravitational lensing work, with the standard survey parameters.  The
instrument properties are taken from Schweitzer et al. 2010
\cite{2010SPIE.7731E..1KS}, Penny et al. 2013
\cite{2013MNRAS.434....2P} and Cropper et al. 2014
\cite{2014SPIE.9143E..0JC}. Since the PSF is proprietary to the Euclid
team, we adopt for simplicity a Gaussian function with full width half
maximum $0\farcs18$.

\item We study the survey mode of the 2.4m WFIRST space mission,
focusing on the 2000 square degree HLS Imaging Survey in the F184W
filter. The instrument/survey properties are based on the
WFIRST-Astrophysics Focused Telescope Assets Final Report (by the
Science Definition Team and WFIRST Project, Spergel et al. 2013
\cite{2013arXiv1305.5422S}) and the WFIRST ETC.
\footnote{\url{http://wfirst-web.ipac.caltech.edu/wfDepc/wfDepc.jsp?etc}}
Since the final PSF was not available at the time of this writing we
adopt a Gaussian function with FWHM $0\farcs15$. As an additional
test, in order to investigate the sensitivity of our estimates to the
choice of the PSF, we repeat the simulations assuming a Moffat (Moffat 1969 
\cite{1969A&A.....3..455M}) PSF, with different shape parameters.

\item The Large Synoptic Survey Telescope (LSST) is designed to
repeatedly image the entire visible Southern sky in six optical and
near infrared bands (out to $\sim$ 1 micron) every few nights for ten
years. We take the survey and instrument properties from Ivezic et
al. 2008 \cite{2008arXiv0805.2366I} and the LSST Science Requirements
Document.\footnote{\url{http://www.lsst.org/files/docs/SRD.pdf}} The
PSF of LSST will vary over time. We adopt a Gaussian PSF with FWHM
$0\farcs7$ as representative of the stacked image quality of LSST in
the i-band, and consider the full depth survey images in our analysis.

\end{enumerate}

We do not consider radio interferometers like ALMA or VLBI, given the
radically different properties of the data, and the need for
additional assumptions relating the optical/IR light to the radio
emission. An assessment of their ability to provide sufficiently
accurate high resolution images for cosmography is left for future
work.


\section{The Lens Sample}

In this section we introduce our sample, taken to be illustrative of the kind of systems to be discovered in the next decade.

\subsection{Sample Selection}

We have chosen four prototypical systems for this exploration,
characterized as {\it faint} or {\it bright} and {\it double} or {\it
quad} depending on the photometry and image configuration. This set of
four main choices covers regimes with different numbers of pixels
above a certain S/N, which in turn depend on the number of images
produced by the lens.

The brighter lenses are selected to be representative of the majority
of currently known lenses, selected from the Sloan Digital Sky
Survey. For reference, the current largest quasar lens sample
comes from the SDSS Quasar Lens Search (SQLS; Oguri et al. 2006
\cite{2006AJ....132..999O}; Inada et al. 2012
\cite{2012AJ....143..119I}).  The well-defined statistical quasar lens
sample of it consists 26 lensed quasars which are selected from SDSS
Data Release 7 (DR7) quasar catalog with Galactic extinction-corrected
i-band magnitudes ($15.0 \leq i \leq19.1$) in the low redshift ($0.6 <
z < 2.2$), adopting morphological (image separations of $1'' < \theta
< 20''$) and color ($i$-band magnitude difference between two images
should be smaller than 1.25 mag) selection algorithms.  Among the 26
lensed quasars, the number of the quad and double images are 4 and 21
respectively, the remaining is a 5-image cluster lens.  Meanwhile, 36
additional lensed quasars are identified from the DR7 quasar catalog,
which contains 4 quad-image systems, 30 double-image systems, the rest
two are 5-image and 3-image cluster systems.

The fainter lenses are selected to be representative of the fainter
systems to be discovered in current and future surveys within the next
decade (Oguri \& Marshall 2010
\cite{2010MNRAS.405.2579O}). We expect that the systems to be used to
measure time delay distances in the next decade will span
approximately the range covered by our sample. Note that some known
lensed quasars are significantly brighter than any of the systems
simulated here (e.g., Patnaik et al. 1992 \cite{1992MNRAS.259P...1P}).
Even though we expect more of those to be discovered in the near
future and they will be useful, they will not represent the majority
of the systems in future large samples, nor present any major
challenge for follow-up, so we do not consider them here.

To remain as realistic as possible, and take advantage of the high
quality information available for the sample, the brighter mocks are
based upon lenses in the Sloan Lens ACS Survey (SLACS; Bolton et
al. 2006
\cite{2006ApJ...638..703B}, Treu et al. 2006
\cite{2006ApJ...640..662T}, Koopmans et al. 2006
\cite{2006ApJ...649..599K}, Gavazzi et al. 2007
\cite{2007ApJ...667..176G}, Bolton et al. 2008a
\cite{2008ApJ...682..964B}, Gavazzi et al. 2008
\cite{2008ApJ...677.1046G}, Bolton et al. 2008b
\cite{2008ApJ...684..248B}, Treu et al. 2009
\cite{2009ApJ...690..670T}, Auger et al. 2009
\cite{2009ApJ...705.1099A}, Auger et al. 2010
\cite{2010ApJ...724..511A}, Newton et al. 2011
\cite{2011ApJ...734..104N}, Shu et al. 2015
\cite{2014arXiv1407.2240S}, Papers I-XII, respectively) and the fainter mocks on the Strong
Lensing Legacy Survey (SL2S; More et al. 2012
\cite{2012ApJ...749...38M}, Gavazzi et al. 2012 \cite{2012ApJ...761..170G},
Ruff et al. 2011 \cite{2011ApJ...727...96R}, Sonnenfeld et al. 2013
\cite{2013ApJ...777...97S}, Sonnenfeld et al. 2013
\cite{2013ApJ...777...98S}, Sonnenfeld et al. 2015
\cite{2015ApJ...800...94S}, Papers I-V, respectively, except for the
first citation) samples.

The sources are set at $z_{\rm s} = 1.071$ and $2.77$ and the
deflectors are set at $z_{\rm d} = 0.351$ and $0.783$, respectively
for the bright and faint systems. The redshifts are taken from the
real systems used as an inspiration for this study, and are well
within the range of the redshifts that we expect to find in future
samples of lensed quasars (e.g. Oguri \& Marshall 2010 
\cite{2010MNRAS.405.2579O}). A different choice within a realistic
range would not have changed any of our conclusions.

\subsection{Parameters of the Sample of Mock Lenses}

The structural parameters of the four systems are listed in Table
\ref{tab:SBProfile} and Table \ref{tab:lenspars}, which are used to describe 
the light model of the host and deflector galaxies and the mass model of the deflector. 
These models will be discussed in more detail in the next section. 
When not explicitly known from existing data,
mock model parameters are assigned via plausibility arguments as
specified below. The source and deflector of the \textit{bright} lens
system configurations are built along SLACS J0330-0020 as modelled by
Bolton et al. 2008a (SLACS V, \cite{2008ApJ...682..964B}), Auger et
al. 2009 (SLACS IX, \cite{2009ApJ...705.1099A}), and Newton et
al. 2011 (SLACS XI, \cite{2011ApJ...734..104N}). For the
\textit{faint} lens system configuration, the source and deflector
have parameters from the Sonnenfeld et al. 2013 (SL2S III,
\cite{2013ApJ...777...97S}) model of SL2S J135949+553550
(also, Sonnenfeld, A., 2015, private communication).

Source and deflector magnitudes in the \textit{bright} case for $I$ and
$V$ bands are from Newton et al. 2011 \cite{2011ApJ...734..104N}, for
$K$-band are estimated based on the other colors, and typical spectral
energy distributions, i.e. an early-type galaxy for the deflector and
a star forming galaxy for the source. Meanwhile, source magnitudes in
the \textit{faint} case are arbitrarily set to 25.0 in all
bands. Deflector magnitudes in the \textit{faint} case are not
directly available from the data, so they require some
extrapolations. The value for $I$-band is based on Sonnenfeld et
al. 2013 (SL2S III,
\cite{2013ApJ...777...97S}), while for $K$ and $V$ bands are estimated
with K-corrections from typical spectral energy distributions. Unknown
magnitudes in $K$-band are assigned from $H$-band ones via
$K_{AB}=H_{AB}$. We note that the host galaxies of type-1 AGNs at
these redshifts are generally brighter than the magnitudes adopted
here (Bennert et al. 2011; \cite{2011ApJ...742..107B}), so our
assumptions are quite conservative.

The effective (half-light) radii to be fitted in \textit{bright}
systems are evaluated separately in different bands, to get a precise
estimate of the effective radius and its uncertainties. Differently, a
unique effective radius is asserted across bandpasses for the
\textit{faint} systems. Neglecting color gradients in this case does
not affect the ability to recover the mass density profile in any
significant way
(Sonnenfeld et al. 2013 \cite{2013ApJ...777...97S}).

For both the \textit{faint} and the \textit{bright} systems, the source positions are
assigned so as to map the source in either two or four images. This
choice allows us to explore whether the number of images makes any
substantial difference while keeping the other properties of the
systems fixed.

In order to construct realistic systems for time delay measurements,
point sources must be added to represent the lensed quasars. As
described in the next section this is done in the image plane, in
order to gain computational precision and efficiency. The
magnification at the location of the quasar images is calculated by
solving the lens equation using \texttt{gravlens}
\cite{2011ascl.soft02003K}. The source-plane magnitudes of the point
sources are chosen to be somewhat brighter then the host galaxy, as it
is typically the case for medium luminosity AGN (e.g., Bennert et
al. 2011 \cite{2011ApJ...726...59B}), and have realistic colors for
AGN at the source redshits. In previous work (Peng et al. 2006a
\cite{2006ApJ...640..114P}; Peng et al. 2006b
\cite{2006NewAR..50..689P}; Rusu et al. 2015
\cite{2015arXiv150605147R}), in general the lensed AGNs are found to be  0 to 3 magnitudes
brighter than the host in the bands considered by our study, although
there are a few exceptions where the host is brighter than the AGN.
The adopted magnitudes are listed in Table~\ref{tab:pointsource}.

We note that the \textit{bright} source configuration is too bright to
be practical for TMT, given its sensitivity -- the exposure time
scales as $D^{-4}$, where $D$ is the telescope diameter, for
background limited point source exposures. Realistically, observations
of such bright systems would be completely dominated by overheads
related to target acquisition and will only be carried out in
extremely rare circumstances. Therefore we do not simulate TMT
observations for the bright systems. Even for the fainter system, we
have chosen to make the AGN artificially fainter
($K$=26.0) in order to avoid saturating the central pixels while
imaging deep enough to get a sufficient signal to noise ratio on the
host galaxy. As will be shown below, the exposure time requirements
for TMT are short even in these cases.

\section{Description of the inference process}

In this section, we summarize the process of modeling the mock lens
systems. By comparing the distribution of the ratio between the
inferred values and the input, we obtain the desired estimate of the
statistical precision with which each parameter can be determined. As
motivated in the introduction, we are interested primarily in the mass
density profile slope $\gamma'$. Therefore, we will only show the
statistics for that parameter, even though all the parameters are
varied simultaneously during the inference. We first describe the
image generating process in Section~\ref{ssec:image}, and then we
describe the inference process in Section~\ref{ssec:inf}.

\subsection{Image generating process}
\label{ssec:image}

Before we fit the mock lens system we need to define the models to
describe the source galaxy light, the lens galaxy light, and the lens
mass. We start with the surface brightness distribution of the
source and lens galaxies. The correspondence between image-plane and source-plane
 is given via the deflection angles, which in turn depend on the mass distribution of the lens.
 A PSF is used to convolve the light of lens galaxy, lensed source galaxy and point-source.
 Finally, we introduce counts noise, readout noise, and background noise.
 These steps are described in this subsection.

\subsubsection{Light Model}

We use the S\'ersic profile (S{\'e}rsic 1963 \cite{1963BAAA....6...41S}, 1968 \cite{1968adga.book.....S})
to describe the surface brightness profiles of the sources and the deflectors. The
S\'ersic profile is described by 
\begin{eqnarray}
   \label{eq:Intensity}
   &I(R) = I_{\mathrm{e}} \exp\left[-k\left(\left(\frac{R}{R_{\mathrm{eff}}}\right)^{\frac{1}{\mathrm{n}}}-1\right)\right] ,\\
   &R(x,y,q) = \sqrt{qx^2+y^2/q}.
\end{eqnarray}
The amplitude $I_{\mathrm{e}}$ is the intensity at the effective
radius $R_{\mathrm{eff}}$, the S\'ersic index $n$ controls the degree
of curvature of the radial light profile, the constant $k$ is determined so that
$R_{\mathrm{eff}}$ is the half light radius (Ciotti \& Bertin 1999 \cite{1999A&A...352..447C}),
 and $q$ denotes the axis ratio. Empirically, $n$ increases with galaxy luminosity, most
galaxies being fitted by S\'ersic index in the range $0.6 \le n
\le 10.0$ (Merritt et al. 2006 \cite{2006AJ....132.2685M}). We adopt $n=4$ for the
deflectors, typical of massive early-type galaxies, and $n=4/3$ or
$n=4$ for the sources, consistent with typical observed values for
blue galaxies and AGN hosts. We checked by re-running simulations with
different $n$ for the source that our results do not depend
significantly on this model assumption.

\subsubsection{Mass Model}

The lens mass profile is assigned within a class of power-law models,
\begin{equation}
\Sigma(x,y)=\Sigma_{cr}\frac{3-\gamma^{\prime}}{1+\qm}\left(\frac{\sqrt{ X^{2}+\qm^{-2}Y^{2}}}{R_{\rm E}}\right)^{1-\gamma^{\prime}},
\end{equation}
where $q_\mathrm{m}$ is the axis ratio, $\gamma^{\prime}$ is the
radial power-law slope, the $\left\{X,Y\right\}$ principal axes are
rotated by the lens position angle w.r.to the canonical $x$
(increasing to the West) and $y$ (increasing to the North).  Within
this functional family, the deflections scale as
$R^{2-\gamma^{\prime}}$ with distance from the lens.  The
\textit{critical density} $\Sigma_{cr}$ is defined by
\begin{equation}
\Sigma_{\mathrm{cr}} = \frac{c^{2}D_{\mathrm{s}}}{4\pi G D_{\mathrm{d}} D_{\mathrm{ds}}}
\end{equation}
in terms of the relative distances to the deflector ($D_{\mathrm{d}}$), to the
source ($D_{\mathrm{s}}$), and between the deflector and the source
($D_{\mathrm{ds}}$). The Einstein radius $R_{\rm{E}}$ is chosen such that, in the spherical limit ($\qm=1$),
 it encloses a mean surface density equal to $\Sigma_{cr}.$ This is also the radius of a ring traced by the host of
 the quasar when this is exactly aligned with the lens galaxy.

 Power-law models of elliptical galaxies (Evans 1994 \cite{1994MNRAS.267..333E}) have often been used with
success over the years to model gravitational lenses.
 Fast methods to compute 2D deflections from elliptical power-law profiles have been given by Barkana 1998
\cite{1998ApJ...502..531B}, as a special case of the formalism by Schramm 1990 \cite{1990A&A...231...19S} for homoeoidal profiles.

Lensed image profiles are obtained by inverse ray shooting. First,
each pixel position in the image plane is mapped back to the source
plane via its corresponding deflection angle. Then, we exploit
the fact that surface brightness is preserved by lensing to assign
surface brightnesses from our putative source models to each detector pixel in the image plane.
Finally, we convolve these images with the PSF appropriate for each instrument, yielding
mock observations of the AGN host galaxy Einstein rings.

\subsubsection{Point Source Model}

In order to reduce computation time and complexity, the point source
images are added to the light model by adding a PSF with appropriate
normalization and position directly to the image plane. The
normalization and position of the images for the center of the source
in the source plane are computed for each configuration by solving the
lens equation using \texttt{gravlens} \cite{2011ascl.soft02003K}.

\subsubsection{Image Noise}

\begin{figure}
\begin{center}
\includegraphics[width=1.0\textwidth]{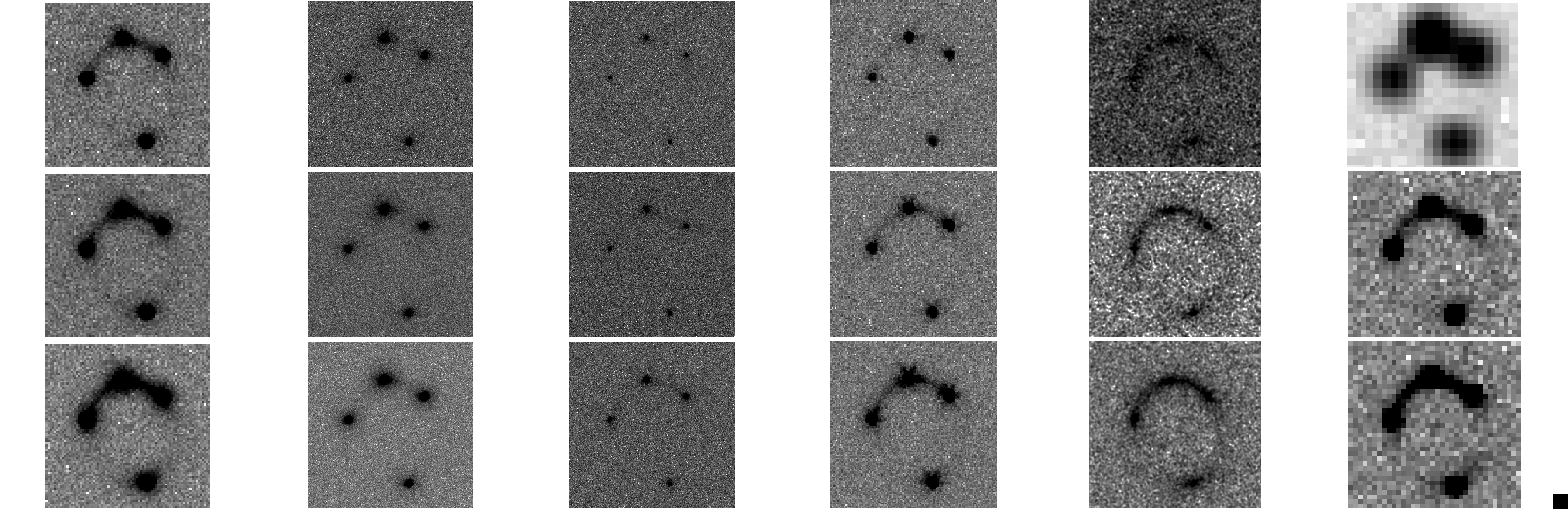}
\end{center}
\caption{Simulations of the fainter lens system with 4 QSO images. The simulated images are all 4$''$ $\times$ 4$''$. The first 5 columns, from left to right, represent HST/ACS, Keck/NIRC2+LGSAO, Keck/NIRC2+NGAO, JWST/NIRCAM, and TMT; from top to bottom, the rows correspond to 1/3 $\times$ ``target'' exposure time, ``target'' exposure time, and 3 $\times$ ``target'' exposure time. ``Target'' exposure time is defined as the exposure time that yields 0.02 precision on the slope of the mass density profile of the mass model $\gamma'$. The sixth column shows simulations of 3 surveys. From top to bottom we show LSST, Euclid, and WFIRST, with the default survey exposure times (4500s, 2360s, 920s, respectively).}
\label{fig:fainter_4QSOimages_montage}
\end{figure}

\begin{figure}
\begin{center}
\includegraphics[width=1.0\textwidth]{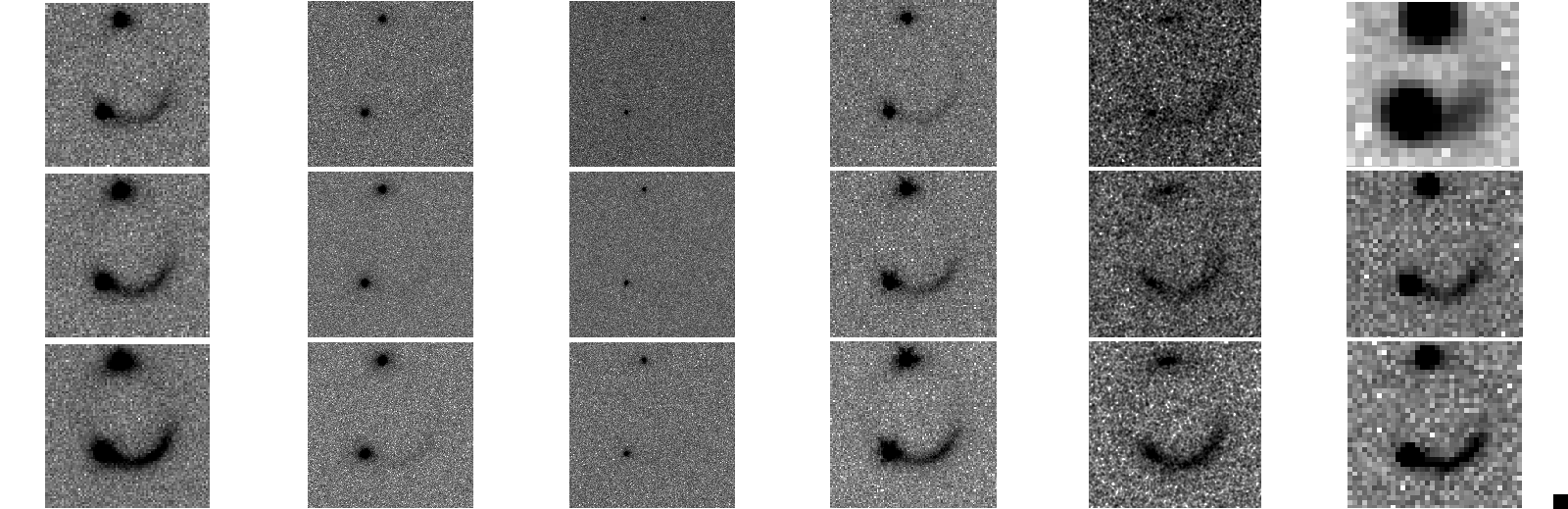}
\end{center}
\caption{Same as Fig. \ref{fig:fainter_4QSOimages_montage}, for the faint double system.}
\label{fig:fainter_2QSOimages_montage}
\end{figure}

\begin{figure}
\begin{center}
\includegraphics[width=1.0\textwidth]{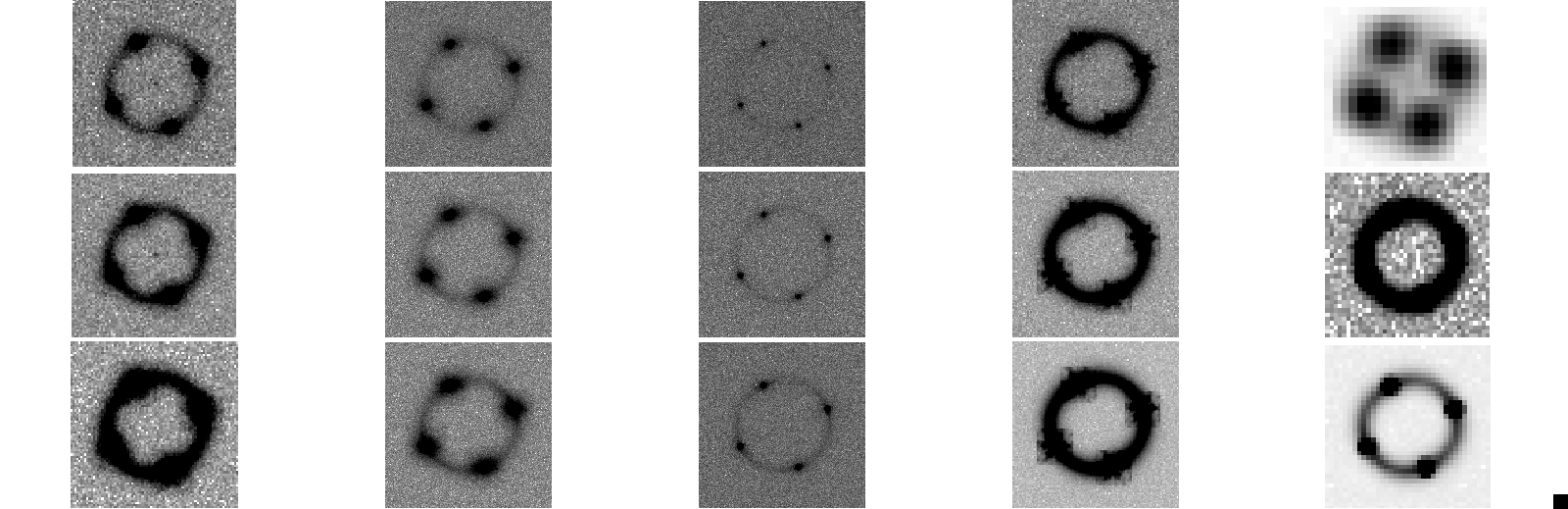}
\end{center}
\caption{Simulations of the bright lens system with 4 QSO images. The simulated images are all 4$''$ $\times$ 4$''$. The first 3 columns, from left to right, represent HST/ACS, Keck/NIRC2+LGSAO, Keck/NIRC2+NGAO; from top to bottom, the rows correspond to 1/3 $\times$ ``target" exposure time, ``target" exposure time, and 3 $\times$ ``target" exposure time. ``Target" exposure time is defined as the exposure time that yields 0.02 precision on the slope of the mass density profile of the mass model $\gamma'$. The fourth column shows JWST simulations with 3 fixed exposure times: 60, 180, 540 seconds, from top to bottom. The fifth column
shows simulations of 3 surveys. From top to bottom we show LSST,
Euclid, and WFIRST, with the default survey exposure times (4500s,
2360s, 920s, respectively). TMT simulations are not shown since the
system is considered too bright to be observed with this telescope in
practice.}
\label{fig:brighter_4QSOimages_montage}
\end{figure}

\begin{figure}
\begin{center}
\includegraphics[width=1.0\textwidth]{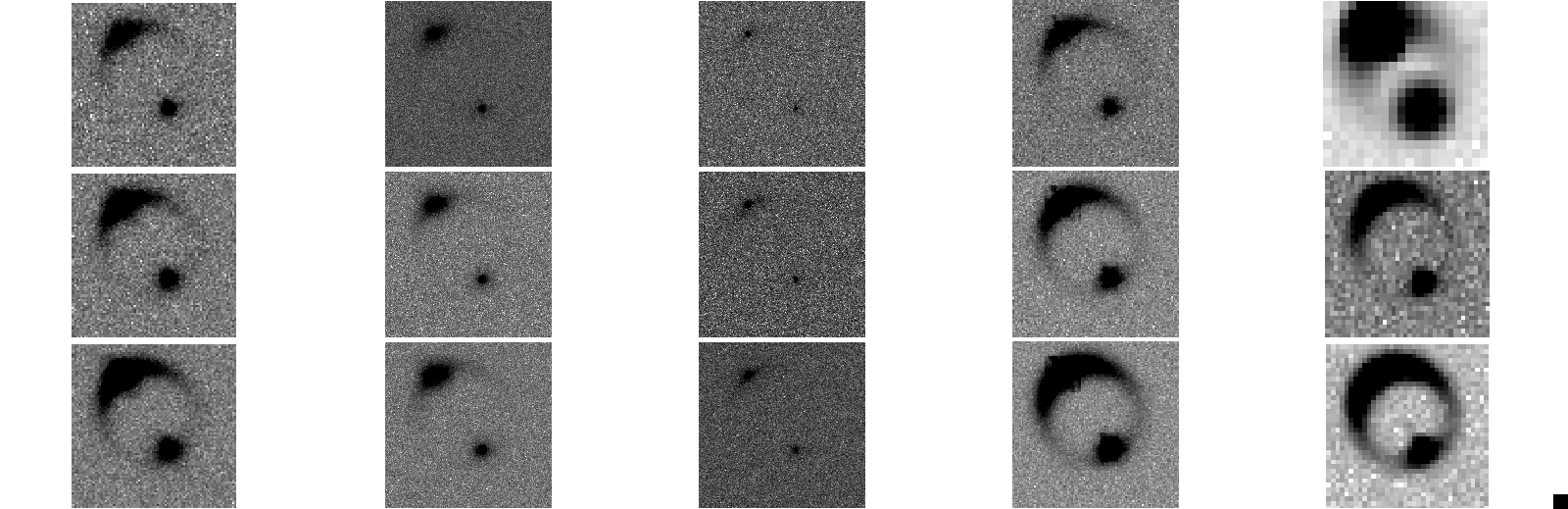}
\end{center}
\caption{Same as Fig. \ref{fig:brighter_4QSOimages_montage}, for the bright double imaged system.}
\label{fig:brighter_2QSOimages_montage}
\end{figure}

In the final step of our simulations we account for the effect of
noise arising from the counts statistics, the background, and detector
read out. As usual, the variance of noise per pixel is given by
\begin{equation}
   \label{eq:noise}
   \mathrm{Var}_{\mathrm{pix}} = \sqrt{Ct+B\cdot t + N_{\mathrm{read}} R^{2}}
\end{equation}
where $C$ is the signal from clean lens system in electrons per second
per pixel, $t$ is the integration time in seconds, $B$
is the sum of the sky background and detector dark current in
electrons per second per pixel, $N_{\mathrm{read}}$ is the number of
detector readouts, and $R$ is the standard deviation of the read noise
in electrons. Realistically, when long integration times are required,
we set the number of readouts by requiring no single exposure be
longer then a maximum time that depends on each instrument
configuration.
Considering the definite properties in different types of instruments,
we pick 1000.0 seconds as the crude maximum exposure time for space
telescopes (HST and JWST), 300.0 seconds for the ground based
telescopes (Keck and TMT) appropriately. Each of the surveys we
model  (Euclid, WFIRST, and LSST) have fixed maximum exposure times
(590 seconds, 184 seconds, and 15~seconds, respectively) which
we use to compute the read noise per exposure. We then emulate  the
stacking of each survey's exposures to reach the approximate expected
total exposure time (2360, 920 and 4500~seconds, for Euclid, WFIRST,
and LSST, respectively).
Examples of simulated images are shown in
Figure~\ref{fig:fainter_4QSOimages_montage}
to~\ref{fig:brighter_2QSOimages_montage}.

\subsection{Inferring the parameters}
\label{ssec:inf}

The same code that is used to generate the mock Einstein Ring images
is also used to fit them. The
modelling procedure follows previous work done with real lens data (e.g., Auger et al. 2011
\cite{2011MNRAS.411L...6A}, Auger et al.~2013 \cite{2013MNRAS.436..503A}, Stark et al.~2013
\cite{2013MNRAS.436.1040S}).
We fit for the following non-linear parameters: position, half light radius, axis
ratio of light, position angle, and S{\'e}rsic index for the
source and deflector brightness model (12 parameters); position, Einstein radius, the
axis ratio of the mass distribution, position angle, and power-law
slope for the deflector mass model (6 parameters); and QSO positions in the image
plane (4 parameters for two-image lenses, 8 parameters for four-image systems). 
The sky level is fixed to the true value in this process. We note in particular
that due to e.g., microlensing, millilensing, and differential reddening, the
positions and fluxes of the QSO images are not tied to the lens model and
they are therefore in effect treated as `foreground' stars. Most priors on the non-linear
parameters are taken to be uniform, but the prior on the position of the deflector
mass is tied to the position of the deflector surface brightness;
the mass is generally close to coincident with the light in galaxy-scale lenses.
The fluxes/amplitudes of each of the surface brightness components
(the source and deflector galaxies and the QSO images) are linear parameters and are determined
by linear inversion for any given set of proposed non-linear paramerters.

The goodness of fit is assessed through the standard
image-plane $\chi^{2},$ by comparing the model and mock data surface
brightness profiles.  Uncertainties are obtained via MCMC exploration
of the posterior, with the fluxes implicitly marginalized over. The likelihood is simply given by
$\mathcal{L}\propto\exp[-\chi^{2}/2]$.

The mock data images for different instruments are all 4$''$ $\times$
4$''$, corresponding to the pixel numbers of 80 $\times$ 80 for HST,
130 $\times$ 130 for JWST, 400 $\times$ 400 for Keck (both LGSAO and
NGAO), 1000 $\times$ 1000 for TMT, 40 $\times$ 40 for Euclid, 36
$\times$ 36 for WFIRST, 20 $\times$ 20 for LSST approximately. The sampling
numbers are fixed at 60000 for each inference progress.
The acceptance ratios remain stable between 20\% $\sim$ 30\%.
Based on the parameters above mentioned, the inference is computationally expensive: typical run times range
between 15 minutes and 4000 minutes per system on a
linux desktop computer. In more specific terms, $\sim$ 100 minutes for HST,
$\sim$ 150 minutes for JWST, $\sim$ 300 minutes for Keck (both LGSAO and NGAO),
$\sim$ 4000 minutes for TMT, $\sim$ 40 minutes for Euclid and WFIRST, $\sim$ 15 minutes for LSST.

\subsection{Exposure times}

The goal of this work is to investigate what combinations of
instrument/telescope configuration and exposure time produce images of
quality sufficient to determine $\gamma'$ with 0.02 precision.

For the telescopes/instruments operated in observatory mode, we are
able to set the exposure time to any desired level. The minimum
exposure time required is defined as the ``target'' exposure time. In
addition, in order to quantify how the precision on $\gamma'$ depends
on exposure time, we also simulated images with 1/3 and 3$\times$ the
target exposure time. This is meant to provide useful guidance for
designing future experiments. However, for some combination of lens
brightness and telescope/instrument configuration, the target exposure
time is too short to be adopted in realistic observations. In
practice, therefore these lenses will either not be observed with this
setup or be observed for longer exposure time. Taking into account
typical overheads for pointing and acquisition we evaluated that the
brighter lenses are too bright to be observed with TMT.  Likewise the
brighter systems are likely too bright to be observed efficiently with
JWST. We still simulated them, but set the minimum exposure time to
60s. Of course in some instances it may be beneficial to obtain longer
exposures with JWST and TMT and thus exquisite data.

For the survey telescopes, we simulated the planned exposure time as
described above, and then  simply considered the specific question of
whether the ensuing data quality is sufficient to meet the requirement
or not.

\section{Results}

\begin{figure}
\begin{center}
\includegraphics[width=0.49\textwidth]{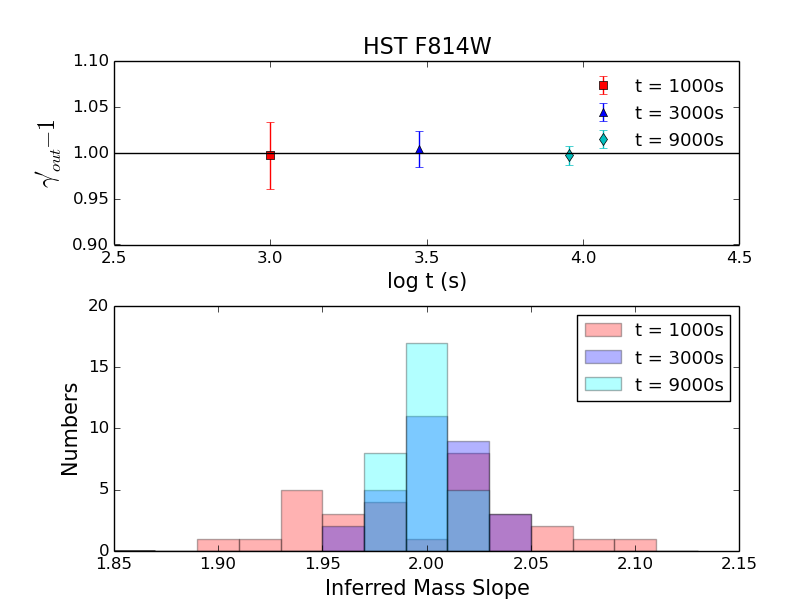}
\includegraphics[width=0.49\textwidth]{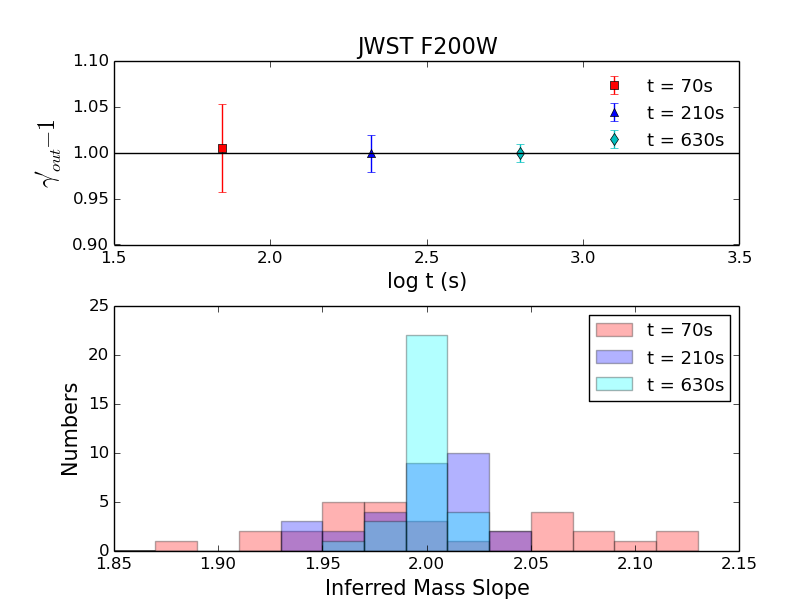} \\
\includegraphics[width=0.49\textwidth]{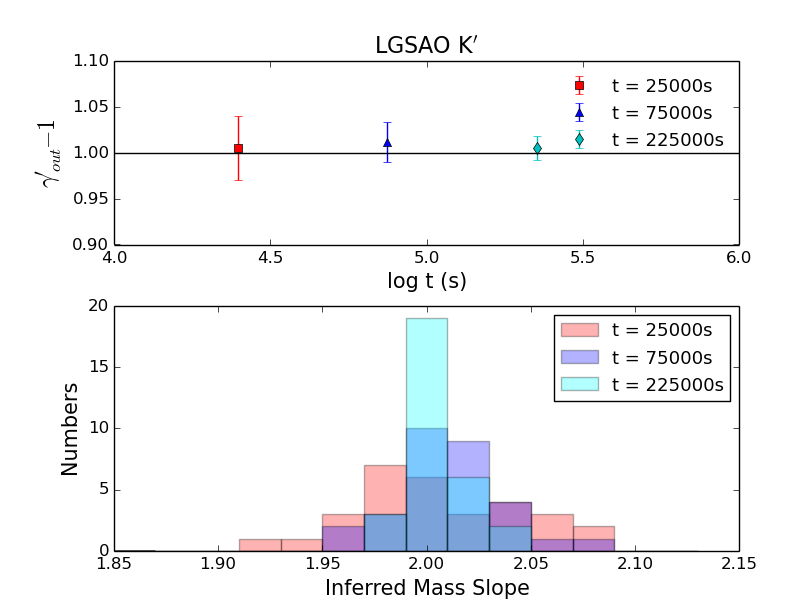}
\includegraphics[width=0.49\textwidth]{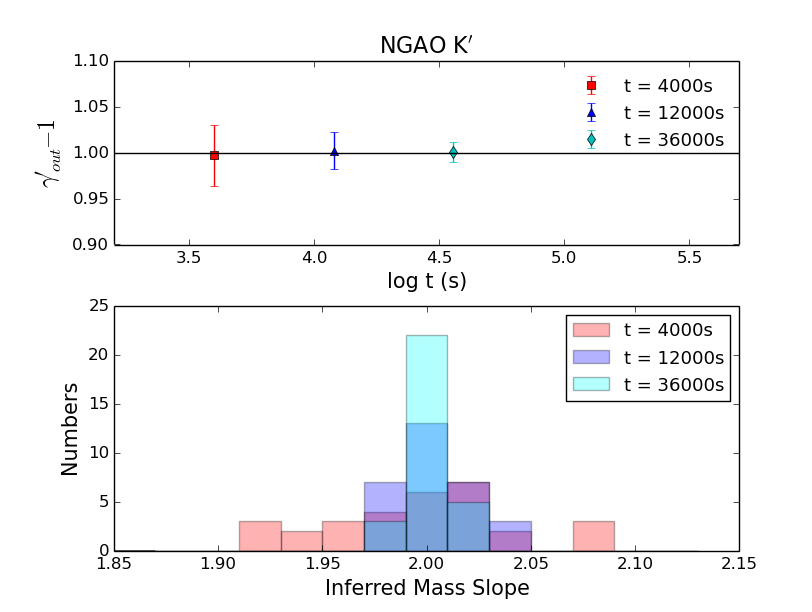} \\
\includegraphics[width=0.49\textwidth]{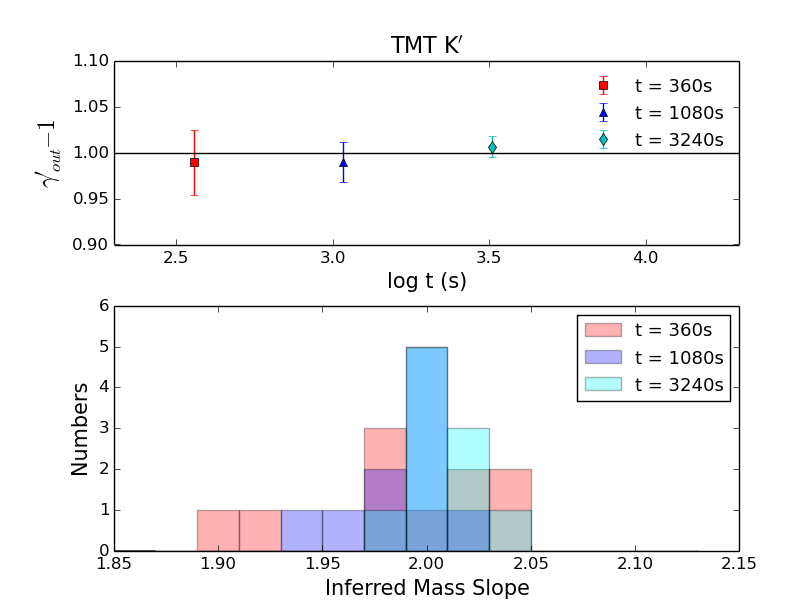}
\includegraphics[width=0.49\textwidth]{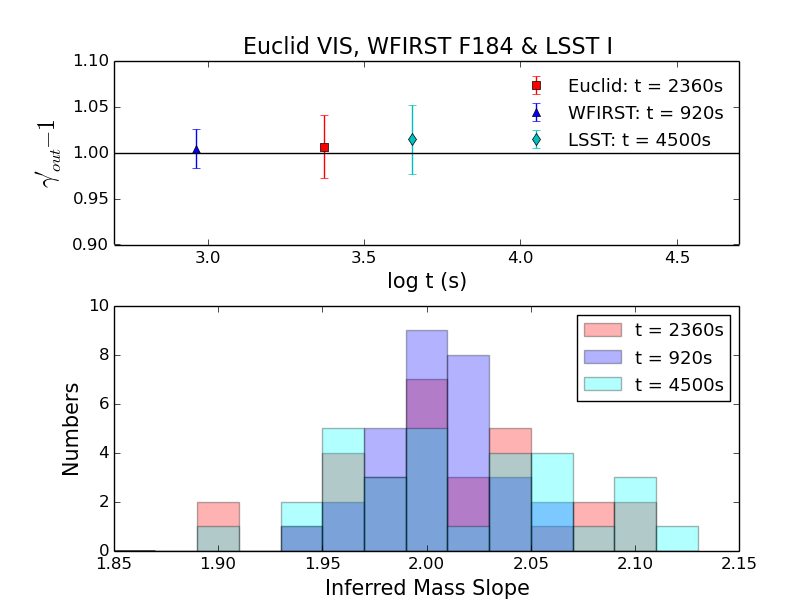}
\end{center}
\caption{Precision on the mass density profile slope $\gamma'$ as a function of exposure time.  This figure shows the results for the fainter lens system with 4 QSO images. The input SIE mass slope is 2. The quantity $\gamma'_{out}$ is the output of the inference process.
The histogram in each panel represents the distribution of inferred
mass slope $\gamma'$ more directly.}
\label{fig:gamma_fainter_4QSOimages}
\end{figure}

\begin{figure}
\begin{center}
\includegraphics[width=0.49\textwidth]{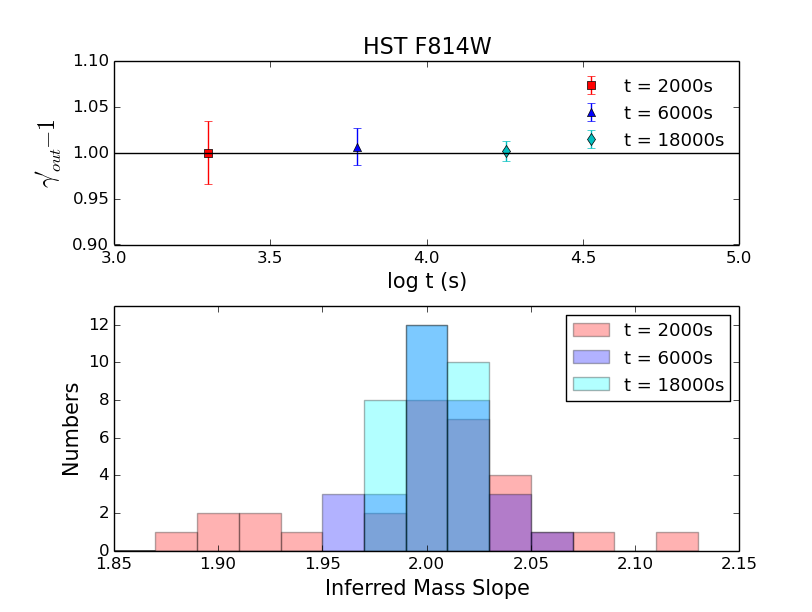}
\includegraphics[width=0.49\textwidth]{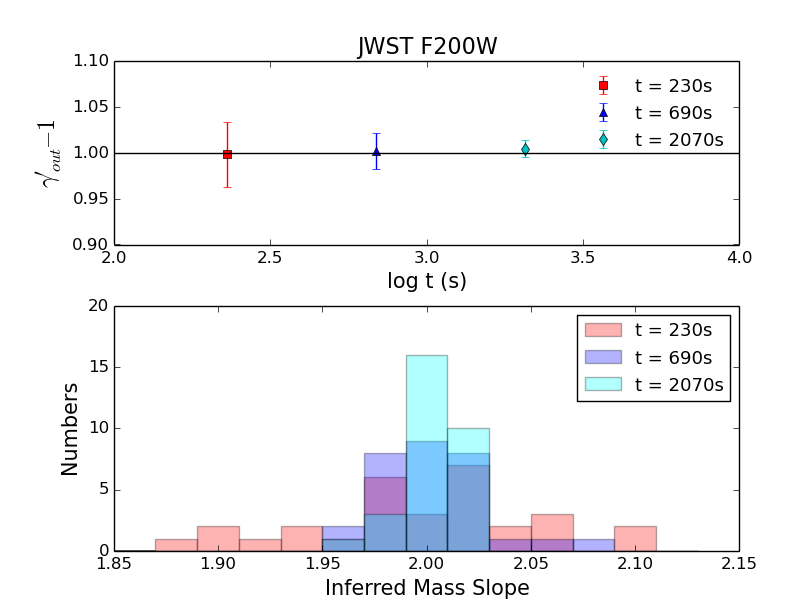} \\
\includegraphics[width=0.49\textwidth]{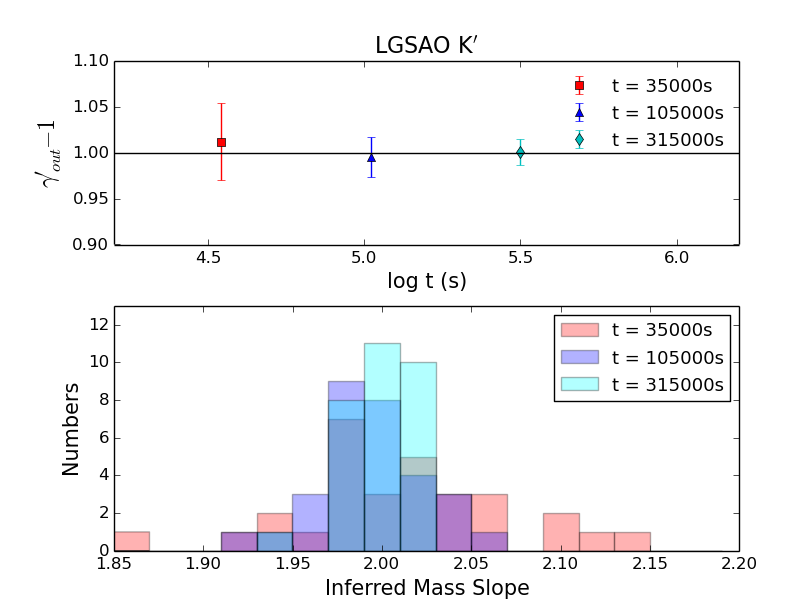}
\includegraphics[width=0.49\textwidth]{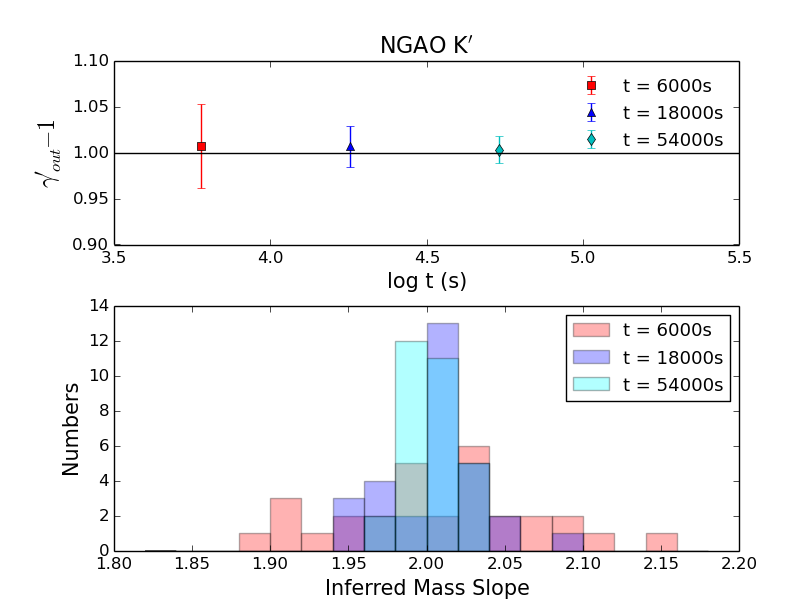} \\
\includegraphics[width=0.49\textwidth]{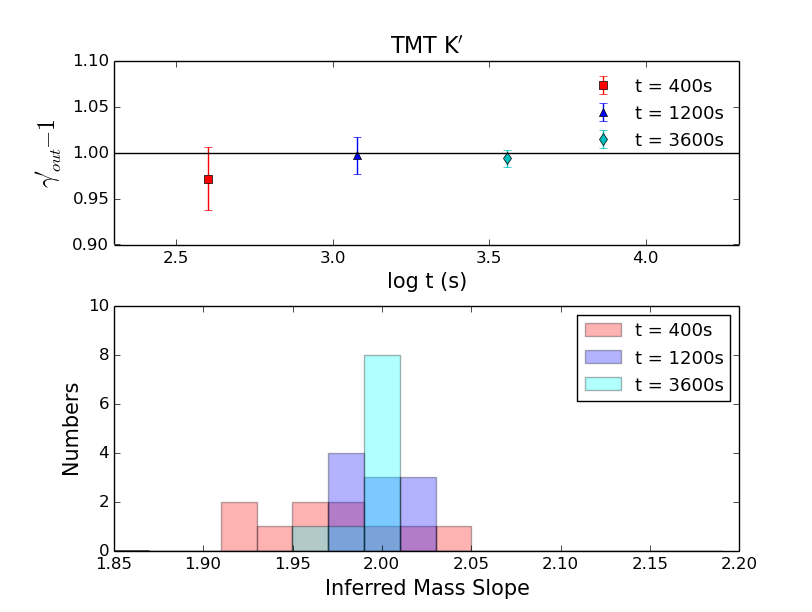}
\includegraphics[width=0.49\textwidth]{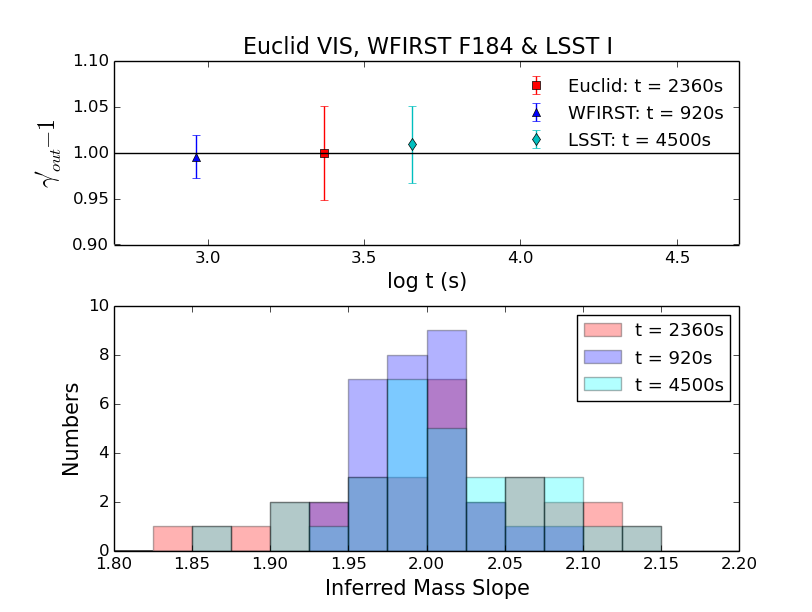}
\end{center}
\caption{Same as Fig.~\ref{fig:gamma_fainter_4QSOimages} for the fainter lens system with 2 images.
\label{fig:gamma_fainter_2QSOimages}}
\end{figure}

\begin{figure}
\begin{center}
\includegraphics[width=0.49\textwidth]{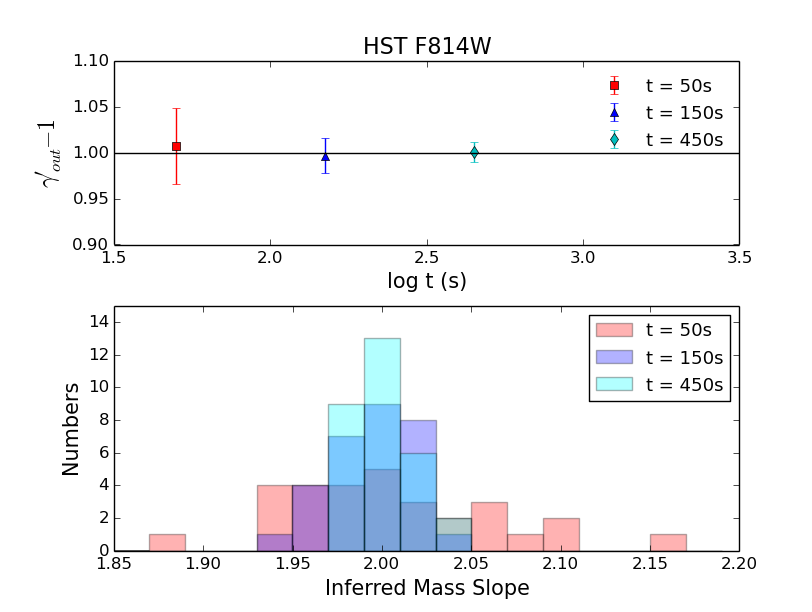}
\includegraphics[width=0.49\textwidth]{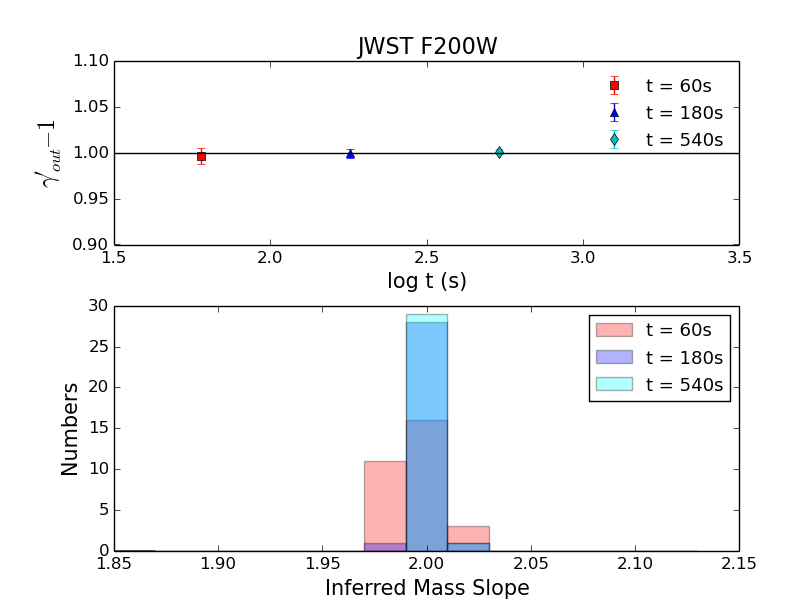} \\
\includegraphics[width=0.49\textwidth]{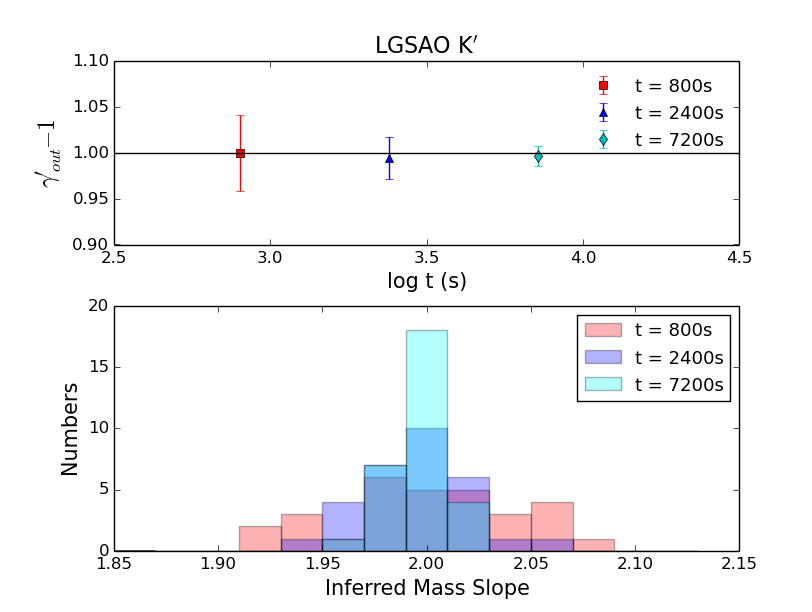}
\includegraphics[width=0.49\textwidth]{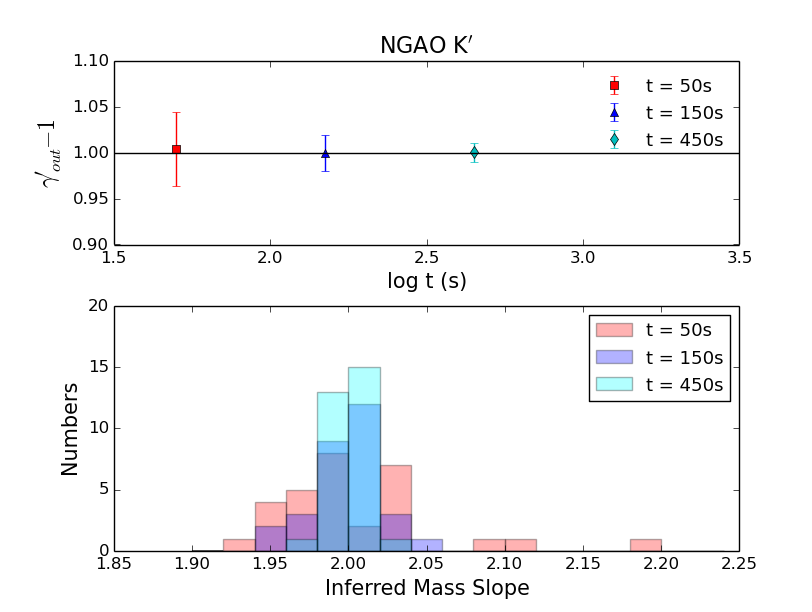} \\
\includegraphics[width=0.49\textwidth]{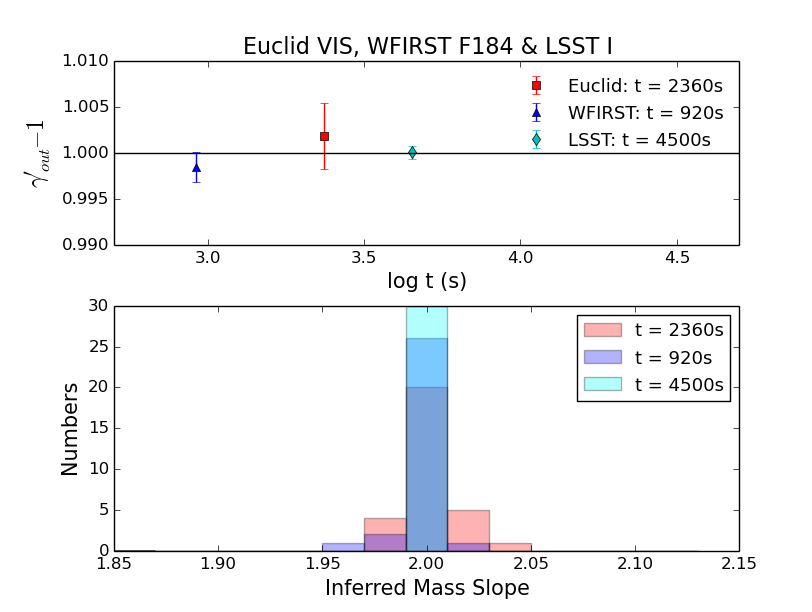}
\end{center}
\caption{Same as Fig.~\ref{fig:gamma_fainter_4QSOimages} for the brighter lens system with 4 images.
\label{fig:gamma_brighter_4QSOimages}}
\end{figure}

\begin{figure}
\begin{center}
\includegraphics[width=0.49\textwidth]{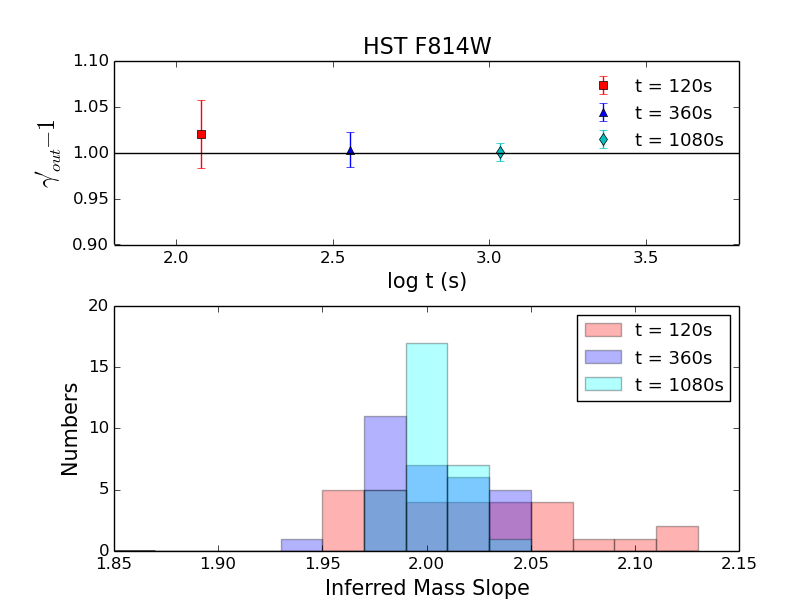}
\includegraphics[width=0.49\textwidth]{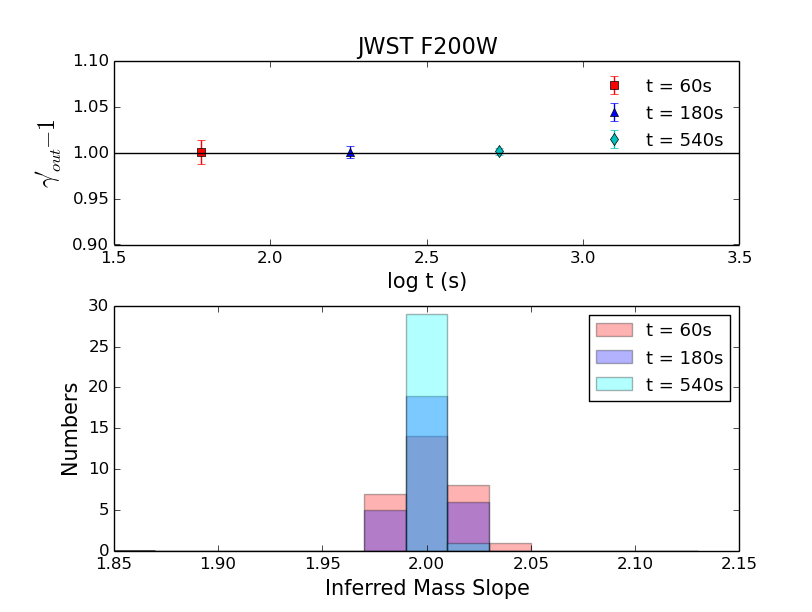} \\
\includegraphics[width=0.49\textwidth]{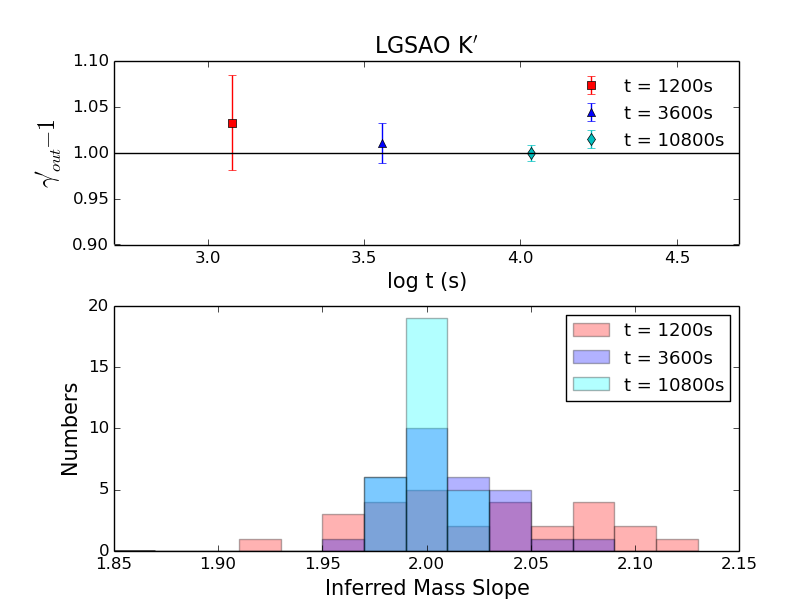}
\includegraphics[width=0.49\textwidth]{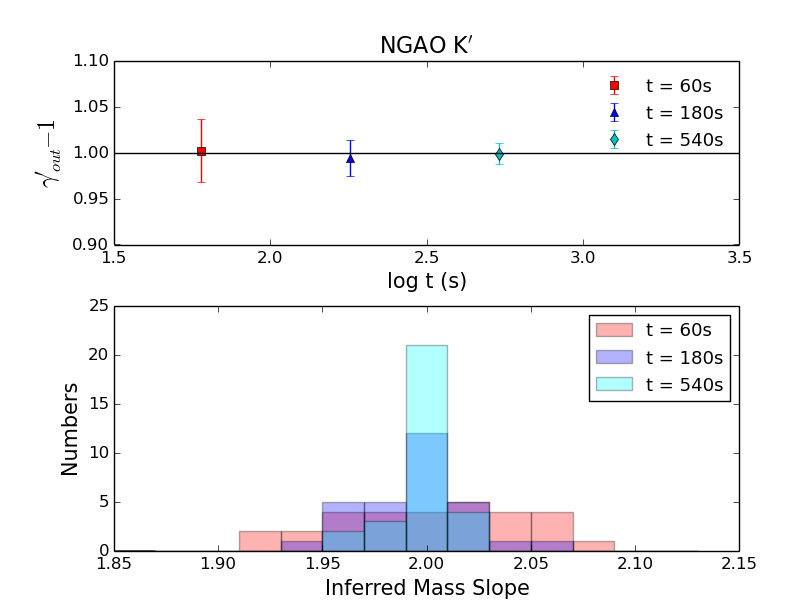} \\
\includegraphics[width=0.49\textwidth]{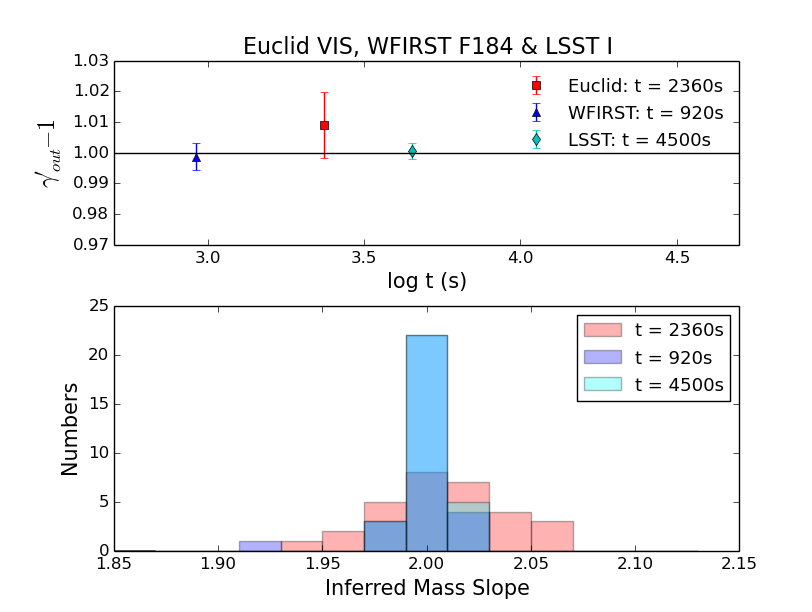}
\end{center}
\caption{Same as Fig.~\ref{fig:gamma_fainter_4QSOimages} for the brighter lens system with 2 images.
\label{fig:gamma_brighter_2QSOimages}}
\end{figure}

Using the methods summarized in the previous section we generate 30
mock systems for most configurations and 10 mock systems for TMT with
the view of saving computing time. The mock systems have the same
true parameters, but different noise realizations. Then we carry out
the full inference for each one using the python Markov Chain Monte
Carlo sampler
\texttt{pymc}.  The results are summarized
in ~Table~\ref{tab:estimated_uncertainties} and ~Figures~\ref{fig:gamma_fainter_4QSOimages}
to~\ref{fig:gamma_brighter_2QSOimages}. Table~\ref{tab:exptimes}
summarizes the ``target'' exposure times for HST, JWST, Keck LGSAO and
NGAO, TMT, i.e. those required in order to reach 0.02 uncertainty on
$\gamma^{\prime}$.  We reiterate that this is is only the random
component of the uncertainty, and thus it represents a lower limit to
the total uncertainty. For this reason we have set a rather stringent
limit of 0.02, in order to leave room for additional uncertainties,
such as, e.g., systematic errors related to PSF reconstruction and
modeling errors.  In practice, our target exposure times should be
considered as minimum exposure times, i.e. necessary and not
sufficient conditions. Assessing the contribution of those additional
sources of uncertainty requires detailed modeling of specific systems
and instruments (e.g., Suyu et al. 2014 \cite{2014ApJ...788L..35S}),
and is left for future work.

First of all, we find that HST can deliver the desired image quality
with exposures of a few kiloseconds at most, consistent with published work
(Suyu et al. 2013 \cite{2013ApJ...766...70S}). Conversely, it seems
that current AO capabilities -- even setting aside the difficulties
associated with the reconstruction of the PSF -- are only sufficient
to study the brighter systems, for reasonable exposure times. The
improvement with NGAO will be substantial, cutting integration times
down to HST-like for both the bright and the faint lenses. In this
case, the higher background than from space is compensated by the
advantage of having smaller detector pixels, and thus better sampling
of the PSF, and a larger telescope aperture. The NGAO performance may
further improve if the thermal background is reduced with respect to
LGSAO as planned and if the instrument has higher throughput than
NIRC2.

For TMT we concluded that 1200s of exposure will be more than
sufficient for all systems considered here and therefore TMT is likely
to be used either in ``snapshot mode'' or to take an image before taking
a spectrum for determining the deflector velocity field and source
redshift, or to follow-up fainter systems, yet to be
discovered. Similary, JWST can obtain images of the desired quality in
just a few minutes, and therefore it is likely to be used mostly to
take images before taking a spectrum or to follow-up fainter systems.

For the surveys, the conclusion is that while the Euclid survey seems
to have insufficient depth and resolution to meet our requirements on
the targets considered here, LSST and WFIRST should be sufficient at
least for the brighter ones. Of course, these surveys will cover vast
fractions of the sky and will therefore provide useful images for all
the brighter systems (Koopmans et al. 2009 \cite{2009astro2010S.159K})
at no additional cost.  Those will be a great complement to the deeper
pointed observation obtained with the other telescopes. WFIRST is
planned to have a component of GO program, so it could meet the
imaging requirements by integrating longer than in the survey mode. In
fact, WFIRST is expected to have imaging performance superior to that
of WFC3 on board HST, which has already been used to perform studies
of time delay lenses, using integration times of a few orbits per
system ($\sim$10ks; HST-GO-12889, PI: Suyu). Amongst the surveys
considered here and modeled with a Gaussian PSF, WFIRST is the one
with most potential for this application. Therefore, it is worth
checking that our conclusion does not depend on the assumption of a
Gaussian PSF. In order to verify this, we have repeated our simulations
adopting a more generic Moffat profile (Moffat 1969 \cite{1969A&A.....3..455M})
\begin{equation}
\label{eq:moffat}
\frac{\beta-1}{\pi \alpha^2}\left[1+\left(\frac{r}{\alpha}\right)^2\right]^{-\beta}
\end{equation}
for the PSF, keeping the FWHM=2$\alpha\sqrt{2^{1/\beta}-1}$ fixed and
varying the shape parameter $\beta$. The results of the simulations
are summarized in Table~\ref{tab:moffat}. The precision on $\gamma'$
decreases only slightly as the wings of PSF become more important
(i.e. smaller $\beta$), showing that our conclusions are robust to the
choice of the PSF shape, as long as that is known.

\section{Summary}

Gravitational time delays are a powerful tool for
cosmography. Transforming measured time delays into time distances
requires modeling the gravitational potential of the main deflector
with sufficient accuracy. Recent work has shown that this is possible
provided that images of sufficient resolution and signal to noise
ratio are available. In anticipation of the hundreds to thousands of
lensed quasars that are going to be discovered in the next decade from
ground based low resolution imaging surveys (Oguri \& Marshall 2010
\cite{2010MNRAS.405.2579O}), we study the requirements for high
resolution imaging follow-up. We consider a range of
instrument/telescope configurations spanning from currently available
systems (HST/ACS and Keck/LGSAO/NIRC2) to planned observatories and
surveys from space (JWST/NIRCAM, Euclid, WFIRST/HLS) and from the
ground (Keck/NGAO/NIRC2, TMT/IRIS). In order to carry out a generic
and systematic comparison across multiple telescopes we simulate four
realistic lens systems, spanning the range of magnitudes and
configuration expected for lenses to be discovered in the next
decade. We also define a single metric, which is the ability to
measure the slope of the mass density profile of an elliptical power
law mass density profile. In order to guarantee that the images be
sufficient to constrain the Fermat potential to a level so that the
random uncertainties are subdominant with respect to other sources of
uncertainty, we consider it necessary for them to contain enough
information to constrain $\gamma'$ with 0.02 precision or better,
which corresponds approximately to a 2\% precision on time delay
distance and hence H$_0$. We find that:

\begin{itemize}

\item Our simulations show that HST can provide sufficient information
with integration times of order 1-10ks, consistent with recent work based
on real data.

\item Keck/LGSAO can provide sufficient information for the brighter
systems to be discovered. The critical issue will be the accuracy with
which the PSF can be reconstructed. Significant effort is under way to
overcome this obstacle (Jolissaint et al. 2014
\cite{2014SPIE.9148E..4SJ}, Ragland et al. 2014
\cite{2014SPIE.9148E..0SR}).

\item The planned Keck/NGAO system will improve the performance of
Keck to make it comparable to that of HST, again provided that the PSF can
be reconstructed with sufficient accuracy.

\item JWST/NIRCAM and TMT/IRIS will be able to deliver the desired
image quality with short exposure times for any lens for which time
delays are reasonably going to be available. Thus they will be ideal
instruments for follow-up of time-delay lenses, especially if short
images are followed by deeper spectroscopic observations to measure
the redshift of the source and the spatially resolved velocity
dispersion (and redshift) of the deflector. For TMT, and other ELTs,
the abilty to reconstruct accurately the PSF of the AO system (Herriot
et al. \cite{2014SPIE.9148E..10H}) will be critical for time delay
cosmography.

\item The survey modes of Euclid/WFIRST/LSST will likely be too
shallow and/or have insufficient resolution except for the very
brightest systems in the sky. Thus Euclid/WFIRST/LSST will make major
contributions in the area of discoveries of these systems, but follow-up
will be necessary for the fainter ones, for example with WFIRST in
general observer mode.

\end{itemize}

We conclude by emphasizing that our study is only concerned with
precision, i.e. the study of random uncertainties. For this reason we
have chosen a target precision of 0.02 on $\gamma'$, that would be a
subdominant contribution to the total error budget for the best
currently studied systems. This is intended to leave room for
additional sources of systematic uncertainty, like for example those
related to the mass-sheet degeneracy. Within instrumental effects, an
important term will stem from uncertainties in the knowledge of the
PSF. Studies based on the comparison of multiple Hubble Space
Telescope optical images have shown that in this case residual PSF
uncertainties do not contribute more than the estimated total
uncertainties. However, the contribution of the PSF to the total error
budget may be more important for ground based Adaptive Optics data and
future telescopes. Significant work is under way to characterize this
source of uncertainty based on existing data (Rusu et al. 2015 
\cite{2015arXiv150605147R}; Suyu et al. 2015, in preparation), even
though we are not aware of a systematic investigation of the
PSF-related uncertainties on the time-delay distance. We plan to carry
out such a systematic investigation in the near future.

\section*{Acknowledgments}

We thank the anonymous referee for helping us improve the paper with a
constructive report.  We are grateful to F.~Courbin, T.~Do,
R.~Gavazzi, C.~Hirata, J.~Larkin, M.~Meneghetti, P.~Wizinowich, and
S.A.~Wright for useful conversations about planned instruments and
telescopes, and P.Schneider for comments on this manuscript. AA, and
TT acknowledge support from NSF grant AST-1450141 ``Collaborative
Research: Accurate cosmology with strong gravitational lens time
delays''. AA, and TT gratefully acknowledge support by the Packard
Foundation through a Packard Research Fellowship to TT.  The work of
PJM was supported by the U.S.  Department of Energy under contract
number DE-AC02-76SF00515.\\

\bibliographystyle{JHEP}
\bibliography{rebuilt_gamma}

\clearpage
\begin{table*}\footnotesize
\begin{center}
\caption{Telescope and Instrument Properties}
\begin{tabular}{lccccccccccccccc|}
\hline \hline
Telescope & Instrument & Filter & Zero Point & Readout Noise & Background Noise & Pixel Size \\
 & & & & (e$^-$/pixel) & (e$^-$/pixel/s) & (arcsec) \\
\hline
  HST    & ACS    &   F814W     &  25.94  & 4.20   &    0.11    &     0.050    \\
  JWST   & NIRCAM &   F200W     &  27.85  & 9.00   &    0.20    &     0.032  \\
  LGSAO  & NIRC2  &   $K^\prime$&  28.04  & 5.75   &      26    &     0.010      \\
  NGAO   & NIRC2  &   $K^\prime$&  28.04  & 5.75   &      26    &     0.010     \\
  TMT    & IRIS   &   $K^\prime$&  31.10  & 2.00   &      21    &     0.004  \\
  Euclid & VIS    &   $r+i+z$   &  25.58  & 4.50   &    0.43    &     0.100   \\
  WFIRST & ---    &   F184      &  26.18  & 5.00   &    0.11    &     0.110   \\
  LSST   & ---    &   $I$       &  28.35  & 5.00   &      68    &     0.200  \\
\hline
\hline
\end{tabular}
\begin{tablenotes}
\item
Observational facilities (telescopes and instruments) considered in this work.
 Zero points are given in the ABmag system. The VIS imager of Euclid spans the whole $r+i+z$ wavelength range. The background and exposure time of Euclid will vary across the sky. Our fiducial values are representative of the average performance. The Keck 10-m telescope is considered both with the current (LGSAO) and next-generation (NGAO) adaptive optics capabilities.\\
\end{tablenotes}
\label{tab:telescopes parameters}
\end{center}
\end{table*}

\begin{table*}\footnotesize
\begin{center}
\caption{Surface Brightness Profile Models}
\begin{tabular}{lcccccccccccccc|}
\hline \hline
 Lens Name & $R_\textrm{eff}$ & $q$ & P.A. & $n$ & $\Delta x$ & $\Delta y$ & $m_{I}$ & $m_{K}$ & $m_{VIS}$ & $m_{H}$ \\
& (arcsec) & & (deg) & & (arcsec) & (arcsec) & & & & \\
\hline\noalign{\smallskip}
\multicolumn{11}{c}{Parameters for the source}
\tabularnewline \hline \noalign{\smallskip}
fainter system$^a$  & 0.23 & 0.92 & 54.0 & 4.0 & 0.0662 & -0.167 & 25.0 & 25.0 & 25.0 &25.0 \\
fainter system$^b$ & 0.23 & 0.92 & 54.0 & 4.0 & 0.008 & 0.298 & 25.0 & 25.0 & 25.0 & 25.0 \\
brighter system$^b$ & 0.12 & 0.77 & 120.0 & 1.33 & -0.195 &  0.34 & 22.73 & 22.0 & 23.46 & 22.0 \\
brighter system$^a$ & 0.12 & 0.77 & 120.0 & 1.33 & 0.01 & -0.005 & 22.73 & 22.0 & 23.46 & 22.0 \\

\hline\noalign{\smallskip}
\multicolumn{11}{c}{Parameters for the deflector}
\tabularnewline \hline \noalign{\smallskip}
fainter system & 1.76 & 0.61 & -9.6 & 4.0 & --- & --- & 20.69 & 19.7 & 21.13 & 19.7 \\
brighter system & 0.91 & 0.81 & 113.2 & 4.0 & --- & --- & 17.99 & 16.5 & 18.84 & 16.5 \\
\hline
\hline
\end{tabular}
\begin{tablenotes}
\item
$R_\textrm{eff}$ is the half light radius. $q$ denotes the axis ratio. P.A. is with respect to the x-axis. The S$\acute{e}$rsic index $n$ controls the degree of curvature of the galaxy light profile. Magnitudes $m$ are given in the ABmag system. The only difference among systems with sources the same brightness is in the source-position, which is set to produce either two or four images of the center.\\
$^a$ This configuration yields 4 QSO images. \\
$^b$ This configuration yields 2 QSO images. \\
\end{tablenotes}
\label{tab:SBProfile}
\end{center}
\end{table*}

\begin{table*}\footnotesize
\begin{center}
\caption{Lens Model Parameters}
\begin{tabular}{lcccccccccccccc|}
\hline \hline
Lens Name & z & $R_\textrm{Ein}$ & q & P.A. & $\gamma'$\\
& & (arcsec) & & (deg) \\
\hline
fainter system & 0.783 & 1.14   &   0.6  &   14.7  &  2.0 \\
brighter system & 0.351 &  1.1    &   0.81  &   113.2  &  2.0 \\
\hline
\hline
\end{tabular}
\begin{tablenotes}
\item
$R_\textrm{Ein}$ is the Einstein radius, according to the definition by Kormann+94.
$q$ denotes the axis ratio of the mass distribution. P.A. is anti-clockwise from the x-axis. The model deflector is a Singular Isothermal Ellipsoid ($\gamma'$=2), which is then fit using power-law models with variable $\gamma'$.\\
\label{tab:lenspars}
\end{tablenotes}
\end{center}
\end{table*}

\begin{table*}\footnotesize
\begin{center}
\caption{Point Source Parameters}
\begin{tabular}{lcccccccccccccc|}
\hline \hline
&  mag(I) &  mag(K) &  mag(VIS) & mag(H) & $\mu$(I) & $\mu$(II) & $\mu$(III) & $\mu$(IV) \\
\hline
fainter system$^a$   &24.0  &   24.0 &  24.0 &  24.0 &  2.84 &  3.75 & 6.59 & 2.78  \\
fainter system$^b$   &24.0  &  24.0  &  24.0 &  24.0 &  2.78 &  2.48 &  ---   &  ---\\
brighter system$^b$ &21.7  &  21.5 &  22.9  & 21.5  & 1.22  & 6.10  & ---   &  ---\\
brighter system$^a$ &21.7  &  21.5 & 22.9  &  21.5  & 6.09  & 8.22 & 6.86  & 7.53\\
\hline
\hline
\end{tabular}
\begin{tablenotes}
\item
Magnitudes (in the ABmag system) and magnifications of multiple images.
 The source magnitude `mag' is given in the source plane so that the observed magnitude in the image plane is given by
  $\mathrm{mag}-2.5\log_{10} \mu$, where $\mu$ is the magnification of each image. We just consider TMT observations for the fainter system, and make the point source artificially fainter ($K$=26.0) in order to avoid saturating.\\
$^a$ This configuration yields 4 QSO images. \\
$^b$ This configuration yields 2 QSO images. \\
\label{tab:pointsource}
\end{tablenotes}
\end{center}
\end{table*}

\begin{table*}\footnotesize
\begin{center}
\caption{Estimated uncertainties on $\gamma'$}
\begin{tabular}{lcccccccccccccc|}
\hline \hline
& & HST & JWST & LGSAO & NGAO & TMT & Euclid & WFIRST & LSST \\
\hline
fainter system$^a$ & 1/3$\times$target  & 0.037 & 0.048 & 0.035 & 0.033 & 0.035 & --- & --- & ---\\
&                        target  & 0.019 & 0.020 & 0.022 & 0.020 & 0.021 & --- & --- & ---\\
&                      3$\times$target  & 0.010 & 0.010 & 0.013 & 0.011 & 0.011 & --- & --- & ---\\
&                       survey  & ---      & ---       & ---      & ---      & ---      & 0.034 & 0.021 & 0.038 \\
\hline
fainter system$^b$ & 1/3$\times$target & 0.034 & 0.035 & 0.042 & 0.046 & 0.034 & --- & --- & ---\\
&                         target & 0.020 & 0.019 & 0.022 & 0.022 & 0.020 & --- & --- & ---\\
&                       3$\times$target & 0.011 & 0.009 & 0.014 & 0.014 & 0.009 & --- & --- & ---\\
&                        survey & ---      & ---      & ---       & ---       & ---      & 0.051 & 0.023 & 0.042\\
\hline
brighter system$^b$ & 1/3$\times$target & 0.037 & 0.013$^*$ & 0.052 & 0.034 & --- & --- & --- & ---\\
&                                       target & 0.019 & 0.006$^*$ & 0.022 & 0.020 & --- & --- & --- & ---\\
&                                      3$\times$target & 0.010 & 0.004$^*$ & 0.009 & 0.011 & --- & --- & --- & ---\\
&                                        survey & ---     & ---             & ---       & ---      & --- & 0.0107 & 0.0044 & 0.0025\\
\hline
brighter system$^a$ & 1/3$\times$target & 0.042 & 0.0087$^*$ & 0.042 & 0.041 & --- & --- & --- & ---\\
&                                       target & 0.019 & 0.0046$^*$ & 0.022 & 0.020 & --- & --- & --- & ---\\
&                                      3$\times$target & 0.011 & 0.0032$^*$ & 0.011 & 0.011 & --- & --- & --- & ---\\
&                                        survey & ---     & ---               & ---       & ---     & ---  & 0.0036 & 0.0016 & 0.0007\\
\hline \hline
\end{tabular}
\begin{tablenotes}
\item
Estimated uncertainties on $\gamma'$ for each instrument.
``target" represents target exposure time which is defined as that reaching 0.02 precision on the slope of the mass density profile of the mass model. $*$ means the estimated uncertainties with artificial exposure time 60s, 180s, and 540s for JWST in brighter systems.\\
$^a$ This configuration yields 4 QSO images. \\
$^b$ This configuration yields 2 QSO images. \\
\label{tab:estimated_uncertainties}
\end{tablenotes}
\end{center}
\end{table*}

\begin{table*}\footnotesize
\begin{center}
\caption{Exposure time requirements}
\begin{tabular}{lccccc|}
\hline \hline
Instrument & \multicolumn{2}{c}{double} & \multicolumn{2}{c}{quad} \\
  & faint  & bright &  faint  &  bright \\
\hline
  HST   & $6\times10^3$s & 360 s  & $3\times10^3$s & 150 s \\
  JWST   & $690$s & 180 s  & $210$s & $<$60 s \\
  Keck (LGSAO)   & $105\times10^3$s & 3600 s  & $75\times10^3$s & 2400 s \\
  Keck (NGAO)   & $18\times10^3$s & 180 s  & $12\times10^3$s & 150 s \\
  TMT   &  1200s  &  ---   &   1080s   &  ---  \\
\hline
\hline
\end{tabular}
\begin{tablenotes}
\item
Exposure times required to attain a 0.02 precision on the inferred $\gamma^\prime.$
 The Euclid, WFIRST and LSST surveys, are not quoted here because their exposure times are fixed. The default exposure times are sufficient to meet the 0.02 requirement in the case of LSST and WFIRST, but not for Euclid.\\
\end{tablenotes}
\label{tab:exptimes}
\end{center}
\end{table*}

\begin{table*}\footnotesize
\begin{center}
\caption{Precision on $\gamma'$ with Gaussian and Moffat PSF for WFIRST}
\begin{tabular}{lcccccccccccccc|}
\hline \hline
Lens Name & Gaussian & Moffat & Moffat & Moffat & Moffat
\\
& & ($\beta$=1.5) & ($\beta$=3.0) & ($\beta$=4.5) & ($\beta$=6.0) \\
\hline
fainter system$^a$ & 0.021 & 0.032 & 0.028 & 0.025 & 0.023 \\
fainter system$^b$ & 0.023 & 0.036 & 0.035 & 0.033 & 0.030 \\
brighter system$^b$ & 0.0044 & 0.0065 & 0.0062 & 0.0048 & 0.0044 \\
brighter system$^a$ & 0.0016 & 0.0022 & 0.0017 & 0.0017 & 0.0015 \\
\hline
\hline
\end{tabular}
\begin{tablenotes}
\item
Precision on the mass density profile slope $\gamma'$ with Gaussian and Moffat PSFs for WFIRST. This table represents the results for fainter and brighter systems with 4 and 2 QSO images. It is shown that as $\beta$ increases the Moffat PSF tends to approximate the core of the Gaussian PSF. \\
$^a$ This configuration yields 4 QSO images. \\
$^b$ This configuration yields 2 QSO images. \\
\label{tab:moffat}
\end{tablenotes}
\end{center}
\end{table*}

\end{document}